\documentstyle[editedvolume,epsf]{crckapb} 

\input psfig.tex

\newcommand{\be}{\begin{equation}}
\newcommand{\ee}{\end{equation}}
\def\lsim{\lower.5ex\hbox{$\; \buildrel < \over \sim \;$}}
\def\gsim{\lower.5ex\hbox{$\; \buildrel > \over \sim \;$}}

\begin{opening}
\title{Accretion Disks Around Black Holes: Twenty Five Years Later}

\author{Sandip K. Chakrabarti}
\institute{S. N. Bose National Centre For Basic Sciences\\
	JD Block, Salt Lake, Sector-III, Calcutta-700091, India\\
        email: chakraba@boson.bose.res.in}

\end{opening}

\runningtitle{Twenty five years of accretion disks}

\begin{document}

\noindent: Appearing in `Observational Evidence for Black Holes in the Universe', published
by Kluwer Academic Publishers, Edited by S.K. Chakrabarti, 1998, page No. 19.


\section{Introduction}

Study of accretion processes onto stars began by the works of Hoyle \& Lyttleton 
(1939), almost sixty years ago. They computed the rate at which pressure-less 
matter would be accreted on a moving star. Subsequently, pressure was included and 
the spherical flow solution was perfected by Bondi (1952). However, emission 
from rapidly infalling matter was not found to be strong enough to explain high luminosities 
of quasars and AGNs. Suggestions to improve the 
luminosity by magnetic dissipation were then put forth (Shvartsman, 1971; Shapiro, 1973ab).
Indeed, efforts to improve luminosity of a spherical flow are on even in recent days 
(Chang \& Ostriker, 1985; Babul, Ostriker \& M\'esz\'aros, 1989; Nobili, Turolla \& Zampieri, 1991).
Meanwhile, possible evidence of disklike structures around one of the binary components  were found
(Kraft, 1963) and some tentative suggestions that matter should accrete in the
form of accretion disks were put forward (Burbidge \& Prendergast, 1968; Lynden-Bell, 1969). 
More quantitative studies were made by Shakura (1972). However, the beginning
of modern accretion disk physics is traditionally attributed to the two classical 
articles, one by Shakura \& Sunyaev (1973, hereafter referred to as SS73) and the 
other by Novikov \& Thorne (1973, hereafter referred to as NT73), both of which
were published exactly twenty five years ago. 

Early history of the development of the accretion disk model has been provided by
Bisnovatyi-Kogan (1998) in this volume and we shall not elaborate on that here. Instead, we shall
dwell on the major developments of this subject, including high points and excitements since 1973.
SS73 and NT73 compute the structure of the accretion disk assuming the angular momentum
distribution is Keplerian, independent of how matter is supplied and independent of the 
nature of the viscous processes. Strictly speaking, no {\it accreting}  disk of {\it finite
temperature} can be perfectly Keplerian at all points. To see this, we have to go back to the
basics. Let's start with a
general set of equations (see Chakrabarti, 1996a, hereafter referred to as C96a)
which the flow satisfies on the equatorial plane around a Schwarzschild black hole (modeled here as
an object around which matter moves in a pseudo-Newtonian potential of Paczynski-Wiita (1980)):

\noindent (a) The radial momentum equation:
$$
v\frac{dv}{dx} + \frac{1}{\rho} \frac{dp}{dx}  + \frac {l^2_{Kep} - l^2}{x^3} = 0
\eqno{(1a)}
$$
\noindent (b) The continuity equation:
$$
\frac{d}{dx}(\Sigma x v) = 0
\eqno{(1b)}
$$
\noindent (c) The azimuthal momentum equation:
$$
v\frac{dl(x)}{dx} - \frac{1}{\Sigma x} \frac{d}{dx} (x^2 W_{x\phi}) = 0
\eqno{(1c)}
$$
\noindent (d) The entropy equation:
$$
\Sigma v T \frac{ds}{dx} = \frac{h(x) v}{\Gamma_3 - 1}(\frac{dp}{dx} -
\Gamma_1 \frac{p}{\rho} \frac{d\rho}{dx}) = Q^+_{mag}+Q^+_{nuc}+Q^+_{vis}-Q^-
$$
$$\ \ \ \ \ \ \ \ \ \ \ \ \  =  Q^+ - g(x, {\dot m}) q^+ = f(\alpha, x, {\dot m}) q^+ .
\eqno{(1d)}
$$
Here, $x$, $v$ and $l$ are the dimensionless distance, velocity and specific angular
momentum (measured in units of $2GM/c^2$, $c$ and $2GM/c$ respectively) from the
black hole. $p$ and $\rho$ are the dimensionless isotropic pressure and
mass density  respectively, $\Sigma$, $W$ and $W_{x\phi}$ are the mass density, 
pressure and viscous stress, integrated over the vertical 
height $h(x)$ (assuming thin disk approximation $h(x)<<x$). $T$ is the 
proton and electron temperature (assuming strong coupling between electrons
and protons, but see, Chakrabarti \& Titarchuk, 1995, hereafter CT95, where 
two temperature flow was studied without the magnetic heating effects).
In the entropy equation, we have included the possibility 
of magnetic heating (due to stochastic field) and nuclear energy release as well.
On the right hand side, we wrote $Q^+$ collectively proportional to the
cooling term for simplicity (purely on dimensional grounds). The quantity
$f$ is a measure of cooling efficiency of the flow.  Also,
$$
\Gamma_3 = 1+\frac{\Gamma_1 - \beta}{4 - 3\beta} ;\ \ \ \Gamma_1 =
\beta + \frac{(4 - 3\beta)^2 (\gamma -1)}{\beta  + 12 (\gamma - 1)(1-\beta)}
\eqno{(2)}
$$
and $\beta(x)$ is the ratio of gas pressure to total (gas plus magnetic
plus radiation) pressure:
$$
\beta(x) = \frac {\rho k T/\mu m_p}{\rho k T/\mu m_p + {\bar a} T^4/3 +
B^2(x)/4\pi}
\eqno{(3)}
$$
$B(x)$ is the strength of magnetic field in the flow, ${\bar a}$ is the Stefan's constant, $k$
is the Boltzmann constant, $\mu$ is the electron number per particle
(and is generally a function of $x$ in case of strong nucleosynthesis
effects), $m_p$ is the mass of the proton.

Now if a disk is strictly Keplerian, $l=l_{Kep}$, and eq. 1a is satisfied only if,
$$
v\frac{dv}{dx} + \frac{1}{\rho} \frac{dp}{dx}=0
$$
at all the points. For a polytropic flow, $p=p(\rho)$, and the integral of the above equation gives,
$$
\frac{1}{2} v^2 -W(p) = W_0
\eqno{(4)}
$$
where, $W_0$ is the value of the potential $W=- \int {\frac{dp}{\rho}}$ at $v=0$ surface.
Now, since the potential must be negative for a bound flow, we see that above equation 
cannot be satisfied unless $W=0=v$ everywhere, i.e., when the flow is strictly non-accreting.

\subsection{Two Corrections of Standard Keplerian Disk Model}

Before we go into the more general state-of-the-art advective disk model in next Section, we wish to
discuss about two corrections to the `standard' notion of a Keplerian disk. First one concerns the angular momentum
distribution equation. Novikov \& Thorne (1973) and all the papers which followed it
too closely (recent one being, Lasota, 1994 which has many other errors) 
used $\tilde{l}= u_\phi$ to be the conserved specific angular momentum. This is not true.
For a fluid with specific enthalpy $h$, the conserved specific angular momentum is $l=h u_\phi$.
This correction in the angular momentum equation was first introduced in Chakrabarti (1996b, 1996c; hereafter referred 
to as C96b, and C96c respectively). See, eq. (8) of C96b.

The second major correction is done in the expression of viscous stress $W_{x \phi}=-\alpha p$. Normally,
in a strictly Keplerian disk where radial velocity is negligible, this form is alright. But when the
inertial pressure or ram pressure $\rho v^2$ is significant, then one has to add this on the right hand side,
$$
W_{x \phi} = -\alpha_\pi (p+\rho v^2)=-\alpha_\pi \Pi.
$$
The effect of $\rho v^2$ is not just cosmetic or rescaling of $\alpha$, but it has a deeper significance. $\Pi$ is conserved
across a discontinuity, or shocks, for instance. Thus, with this definition, viscous stress would also
remain continuous across the shock. This ensures that no undue transport of angular momentum takes place in the disk. In a smooth
flow this definition should be used, particularly, when infall velocity is significant as in an advective disk.

\section {Need and Attempts for a Disk Model Alternative to Keplerian }

In a black hole accretion, the specific binding energy on an equatorial
plane ($\theta=\pi/2$) is give by (C96b, C96c),
$$
u_t= - \left[\frac{\Delta}{(1-V^2)(1-\Omega l)(g_{\phi\phi}+l g_{t\phi})}\right]^{1/2} ,
\eqno{(5)}
$$
where, $g_{\mu\nu}$ is the metric coefficient, and,
$$
\Delta= r^2 - 2 r + a^2; \ \ \ \ \ \omega = \frac{2 a r}{A} ; \ \ \ \ \ A= r^4 + r^2 a^2 + 2 r a^2
$$
with $a$ is the Kerr parameter and  $r$ is the radial coordinate. $V$ is the radial velocity measured in the
rotating frame, $l$ is the conserved specific angular momentum and $\Omega$ is the 
angular velocity of the orbiting matter. On the horizon, $\Delta=0$, hence
for a flow with finite binding energy on the horizon, $V=1$. Since for causality, the speed of sound, $a_s <1$, flow must be 
supersonic on the horizon (Chakrabarti, 1990; C96b). When a flow, strictly Keplerian at a large distance,
enters through the horizon with $V=1$, it must cross a sonic point at an intermediate distance. The flow must
start having large radial velocities as it approaches the horizon, i.e., the flow must be advective. 
It can carry energy, entropy along with matter when advecting.
Abramowicz \& Zurek (1981) computes the sound speed at the sonic point to be,
$$
a_s^2 = \frac{1}{x_c^2} [l_k^2(x_c) - l^2]
\eqno{(6)}
$$
and concludes that the flow must be sub-Keplerian at that point, since, by definition, $a_s^2>0$. However,
this is not generally true when viscous flow with cooling, heating, nuclear energy release etc. are
considered (C96a). The flow could be Keplerian or even super-Keplerian at the sonic point. What is
definitely true, however, is that irrespective of any heating or cooling effects, the flow must be
{\it sub-Keplerian} as well as supersonic at the horizon.

This necessitates the study of non-Keplerian flows more seriously. Indeed, even the observations from {\it Uhuru}
back in the seventies of Cyg X-1 required a non-Keplerian component. This component 
was thought to be  Compton clouds in models of Zdziarski (1986) or a coronal layers with possible holes
in two-phase models of Haardt \& Maraschi (1991). 
Recently, Chakrabarti (1995), CT95, Chakrabarti (1997a; hereafter C97a) looked into this
problem including advection and bulk motion and the theoretical predictions of the advective flows were fully put to test.

\subsection{Attempts to Non-Keplerian Flow Study: Thick Accretion Disks}

The pressure effect ($\frac{1}{\rho} \frac{dp}{dx}$ of eq. 1a)
was first included by Paczy\'nski and collaborators and others
in a series of papers (Abramowicz et al. 1978; Ko\'zlowski  et al. 1978; Jaroszy\'nski et al,
1980; Paczy\'nski et al., 1980; Chakrabarti, 1985). Pressure from radiation causes the 
flow to be non-Keplerian, and fattens the disk geometrically. Matter with angular momentum refuses to
come closer to the axis of the disk, and a funnel wall is also produced. All these were very exciting,
since the origin and formation of jets was a real problem, and the funnel wall, with its super-Eddington
luminosity seems to be helpful to push matter out  along the axis.  Meanwhile, Rees et al. (1982) 
also suggested (albeit using qualitative considerations) that the thick accretion disks are possible for very low 
accretion rate, since the gas would remain very hot with virial temperature $T \sim GMm_p/{rk}$ 
(where, $k$ is the Boltzman constant). However, these disks did not have any radial velocity
and early attempts to include this (see next Section) were not very successful. Normal proceedings was
perturbed by the so-called Papaloizou-Pringle (1985) instability  which was not found to be so 
fatal after all (e.g., Kojima, 1986). After two dimensional advective disks came about 
these thick disks are no longer studied in isolation, since
in presence of centrifugal barrier, these `classical' thick disks are special cases of advective disks.

\subsection{Advection Models: Ups and Downs}

Meanwhile, development was going on to include advection term ($v\frac{dv}{dx}$ of eq. 1a) as well.
A large number of workers, since 1980s realized that accretion disks, both thin and thick, 
need a face-lifting by addition of radial velocity. However, earlier works 
had a partial success. Works of Liang \& Thomson (1980), Abramowicz \& Zurek (1981)
Paczy\'nski \& Bisnovatyi-Kogan (1982), Paczy\'nski \& Muchotrzeb (1982), Muchotrzeb (1983) etc. deserve
some mention in this respect.  Several conclusions drawn in these works were incorrect, specially the 
existence of a special $\alpha$ parameter in Muchotrzeb (1983) was contested by Abramowicz \& Kato (1989) as an
artifact of finite distance of the boundary. Secondly,
six B parameters describing vertical averaging was narrowed down in slim disk model
of Abramowicz et al. (1988). However, while correcting a set of errors new errors and wrong
concepts were introduced. (This is quite normal in the developmental phase of a subject.)
One conclusion which excited the community temporarily was that the black holes 
should allow multiple solutions, just because there are multiple sonic points (Abramowicz \& Zurek, 1981). 
Today we know that
the entropies of the flow at these two points are completely different, and this
usually means two completely different flows pass through two different sonic points. In fact, by topological
reason one of the solutions cannot enter into the black hole at all. On the other hand, these two flows
with two different entropies could be connected by standing shocks to obtain a unique steady solution (C96a, Chakrabarti, 1996d;
hereafter C96d) since irreversible increase of entropy takes place at the shock. 
Matsumoto et al. (1984) tried to repair the 
inner edge of a strictly Keplerian flow by introducing transonic flows with nodal type sonic points. Similarly, both
the methods and the global solutions of slim disks (including later solutions named advection
dominated flows which follow this approach) are not correct. For instance,
by choosing arbitrary initial velocity, sound speed, angular momentum etc., the solutions obtained
were found to be globally  incorrect. Fig. 3 of Abramowicz et al. 1988 suggests that 
(a) angular momentum should deviate away from a Keplerian disk as it approaches a Keplerian disk
(for $r_{out}=10^2$ case), and (b) the flow with accretion rate $50$ times the {\it critical rate},
i.e., $800$ times the Eddington rate was found to be deviating from a Keplerian disk
at $r_{out}=10^5 R_g$, where $R_g = 2GM/c^2$, the Schwarzschild radius. Today, we know that both these
solutions are wrong.  Both of these errors were coming from the initial condition. All the 
errors in slim disk approach have propagated into its recent re-incarnation 
of advection dominated flows (ADAF) which are supposed to deviate from a Keplerian disk by efficient evaporation
typically at a million 
Schwarzschild radius (Narayan \& Yi, 1994, 1995 and Narayan 1997). 
It is easily shown that this is impossible especially for a high viscosity flow which ADAF uses.

\section{Fundamentals of Advective disks}

An advective disk is the one which advects, or carry `something', namely, mass, entropy, energy  etc.
Since this fundamentally means that radial velocity must be present, I define advective disks as those
which have finite radial velocity which may even reach the velocity of light (e.g., on the horizon).
Whether they actually advect energy or not will depend on the accretion rate and viscosity which 
in turn decide the cooling and heating efficiencies. These disks are the most general which are studied
so far. If slim disks (high accretion rate, optically thick) or ADAF (low accretion rate and high viscosity, with very
low radiative efficiency) solution really exists they would automatically come out of the advective
disk solutions. For a black hole accretion, advective disks are the same as the transonic
disks. For a neutron star accretion this is not necessarily so, as the
neutron star accretion could be completely subsonic as well. Some of the recent reviews of the
advective disks are in C96d and Chakrabarti (1998).

\begin{figure}
\vbox{
\vskip +0.0cm
\hskip 0.0cm
\centerline{
\psfig{figure=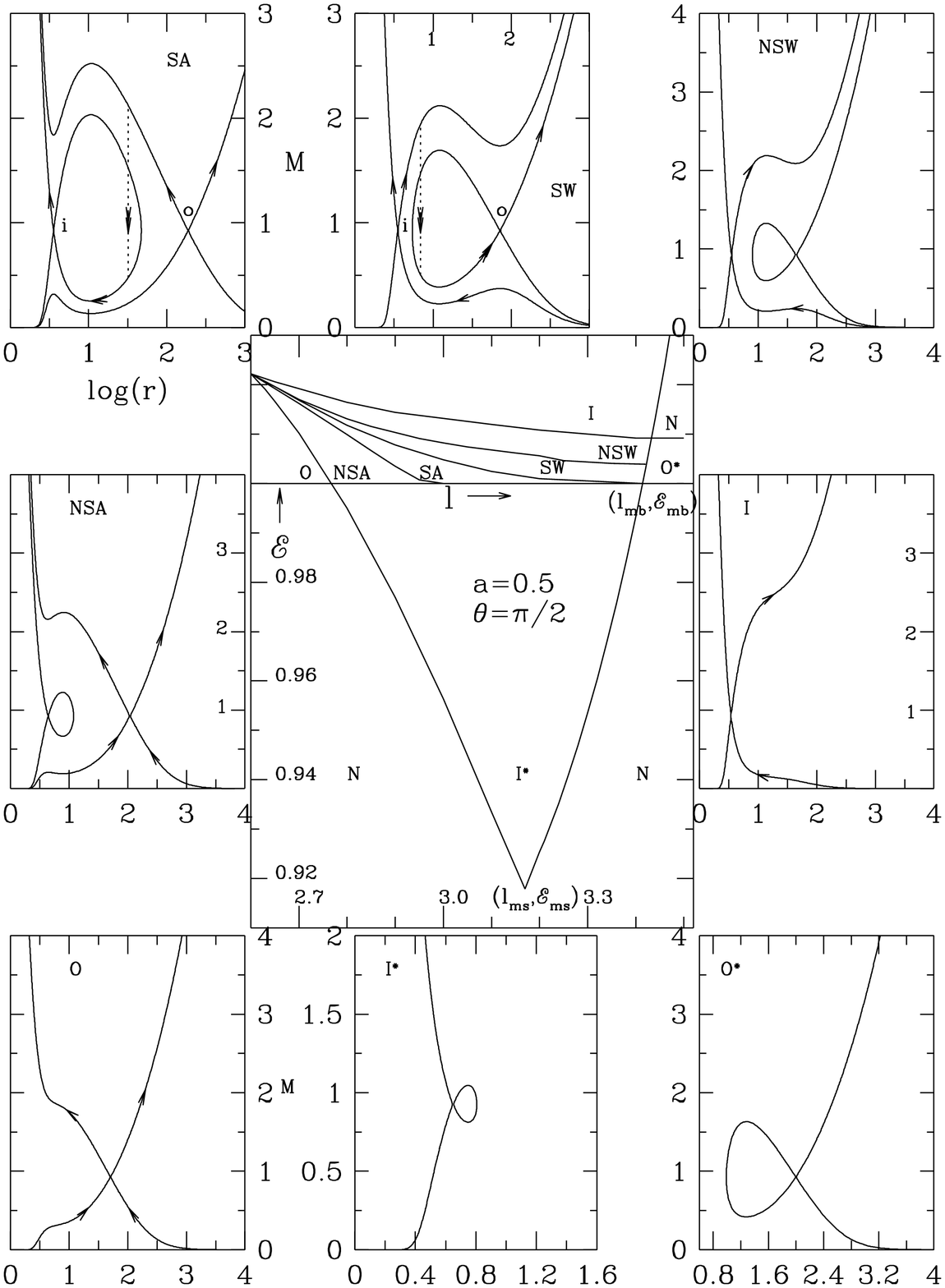,height=12truecm,width=12truecm,angle=0}}}
\vspace{0.5cm}
\noindent {\small {\bf Fig. 1} : Classification of the parameter space (central box)
in the energy-angular momentum plane in terms of various topology of the
black hole accretion. Eight surrounding boxes show the solutions from
each of the independent regions of the parameter space. Each small
box shows Mach number $M$ against the logarithmic radial distance $r$
(measured in units of $GM_{BH}/c^2$)
Contours are of constant entropy accretion rate
${\dot{\cal M}}$. Similar classification is possible for all adiabatic index
$\gamma <1.5$. For $\gamma >1.5$, only the inner sonic point is possible
other than an unphysical `O' type point [C96b]. }  
\end{figure}
Before the full fledge advective flow is presented, consider only the inviscid, constant 
angular momentum thin flow in a Kerr geometry. Figure. 1, taken from C96b 
(also see, Chakrabarti, 1989; hereafter C89) shows all possible solutions and non-solutions. Mach number is
plotted against the logarithmic radial distance (measured in units of $GM/c^2$). If the matter is bound, i.e., when the
specific energy ${\cal E}$ is less than the rest mass of the flow $c^2$ (where $c$ is the
velocity of light, which is taken to be unity here), then there is {\it no} complete solution which becomes transonic 
(see, region ${\cal E} \leq 1$ in the Figure). See, Ryu (this volume) for further
discussion on this classification. Thus a cool, Keplerian flow which is bound
everywhere, does not have a way to enter into a black hole. Only if enough
viscosity is present, then closed topology of $I^*$ opens up (Fig. 3 below) and the originally
Keplerian flow
enters through the inner sonic point. However, these transonic flows would join with the Keplerian
disk {\it very close} to the black hole (roughly around the inner sonic point). These 
are not ADAF solutions. In ADAF (Narayan \& Yi, 1994, 1995) 
the flow starts to deviate from Keplerian disk (due to evaporation) from a million Schwarzschild radii.

When the flow is away from the equatorial plane, or energized by magnetic flares 
or other coronal effects, or in the extreme case, when the Keplerian disk itself 
is {\it very hot}, so that the specific energy is greater than $1$ (rest mass), flow would 
deviate from a Keplerian disk and pass through outer (O, NSA), or inner (I, NSW), or both 
(NS, SW) sonic points. Flow may (SA, SW) or may not (NSA, NSW) have
a standing shock in the flow. Only very high energy flows with weak viscosity (I)
or very low energy flows with high viscosity ($I^*$) pass through the inner sonic 
point. This figure is drawn for a flow in vertical equilibrium and for adiabatic 
index $\gamma=4/3$. It is to be noted that only one sonic point would be 
present if the polytropic index is greater than $\gamma=1.5$ so the subdivision 
of the parameter space would look different and the question of shocks do not arise
(Note a typographical error in C97 where it was mistakenly  stated that $\gamma <1.5$ would not have shocks.).

This classification of solution, though done for inviscid flows, is the backbone of 
the advective disk physics. Viscosity and cooling modify these solutions by changing 
the topologies in a very predictable way (see Fig. 3 below). But close to a 
black hole, where the infall timescale $r/v(r)$ is short compared to the viscous timescale
(unless $\alpha \geq 1$) viscosity does not do much. Angular momentum remains roughly constant
(and therefore behaves like an inviscid flow in some sense)
in the last few to a couple of tens of Schwarzschild radii. Constant angular momentum
flow introduces large centrifugal force which forms a dense region around a black hole (CENBOL).
This is the centrifugal pressure supported boundary layer of the black hole. If the viscosity is small, this barrier
is prominent, and even standing shocks may form, but when the viscosity is  very large,
the barrier may disappear and matter virtually falls freely near the horizon. This unique region
produces the power-law hard tail in the soft states of black holes through
{\it bulk motion Comptonization} as described in CT95.

\begin{figure}
\vbox{
\vskip 0.5cm
\hskip 0.0cm
\centerline{
\psfig{figure=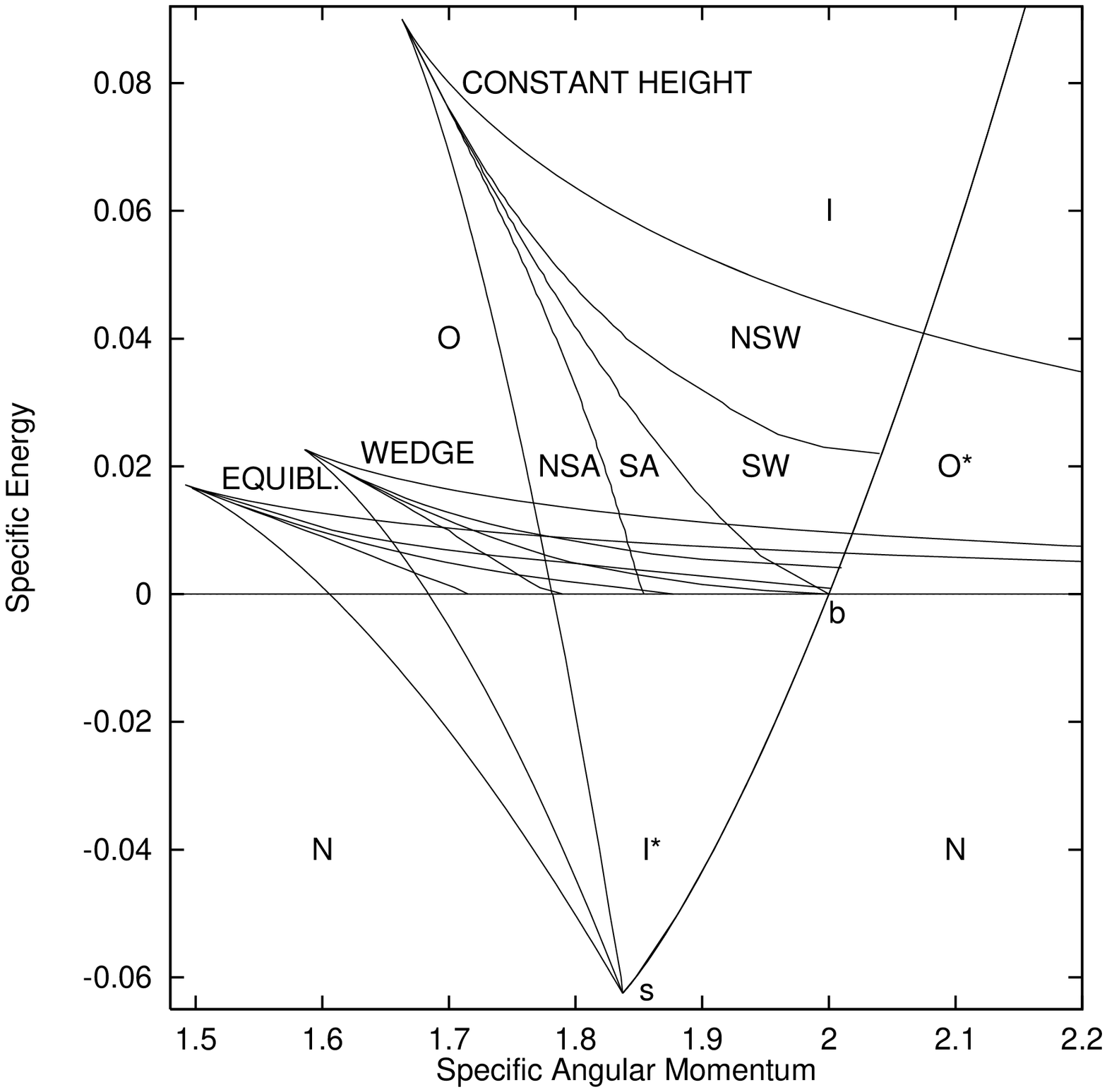,height=7truecm,width=8truecm,angle=-90}}}
\vspace{+0.1cm}
\noindent {\small {\bf Fig. 2} :  Same as in the central box  of Fig. 1, but for flows of various
models in pseudo-Newtonian geometry. When the models are changed  as marked, 
the number of regions, and therefore topologies, do not
change. Only the boundaries vary. The points $s$ and $b$ which represent the
marginally stable and marginally bound quantities of the geometry remain invariant under change of models.}  
\end{figure}
The classification presented in Fig. 1 is generic. The total number of 
topological variation of the solutions does not depend on any cosmetic changes of the model,
such as vertical averaging. Instead, the boundaries of the region will vary.
Figure 2 shows the classification in a Schwarzschild geometry for 
a (a) thin disk with constant height, (b) conical wedge flow and (c) a flow 
in vertical equilibrium. Note that in each of the models the point  $s$
and $b$ are fixed. They represent the marginally stable and marginally bound
quantities and are functions of the geometry only, and therefore model
independent. However, since it is a pseudo-Newtonian model, the point $s$
occurs at ${\cal E}=-0.0625$ rather than at $-0.057$, valid for Schwarzschild geometry.
Note also that since it is a Newtonian computation, rest mass has been subtracted from the
specific energy. Thus, cool Keplerian flows have negative specific energy in this notation.

Because the flow is inviscid, the angular momentum is constant and the flow cannot join 
with a Keplerian disk within a finite distance at all (i.e., joins at infinity!). The energy is
conserved, so the entire energy is advected towards the black hole. In presence of shocks (which form in regions
$SA$ and $SW$ in accretions and winds respectively) entropy can be generated due to
turbulent viscosity (for instance) at the shock and the entropy generated is also
advected away completely (C89, Chakrabarti, 1998). Thus,
a truly advection dominated flow (completely radiatively inefficient) is the most weakly viscous flow
which joins with a Keplerian disk very far away. But, then, the accretion rate should
be excruciatingly low so that the energy remains roughly conserved. 

\begin{figure}
\vbox{
\vskip 0.0cm
\hskip 0.0cm
\centerline{
\psfig{figure=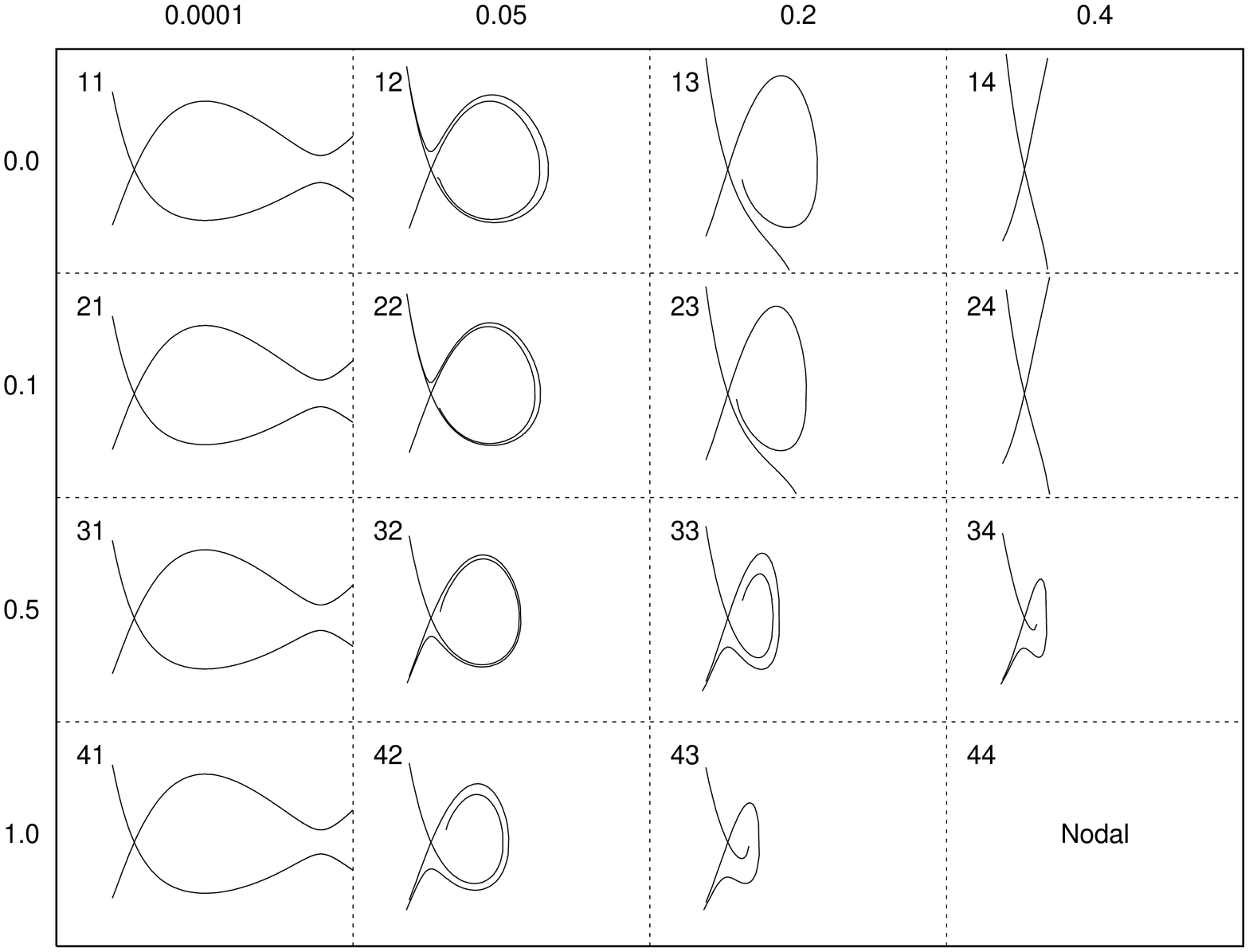,height=8.5truecm,width=12truecm,angle=-90}}}
\vspace{-0.0cm}
\noindent {\small {\bf Fig. 3}: Changes in advective disk topologies when viscosity parameter $\alpha_\Pi$ 
(marked on upper axis) and the cooling factors $f$ (along left axis) are varied. $x_{in}$ and $l_{in}$
are kept fixed. ADAF type solutions ($f \sim 1$) are impossible for high viscosity. Logarithmic
distance along x-axis and Mach number is along y-axis. See text for details.}
\end{figure}
This understanding is fully confirmed when one deals with a general flow as described in C96a.
Figure 3 (Fig. 2a of C96a) shows a class of solutions placed in grids (grid numbers are marked in each box)
of size $0$ to $50 R_g$ along
$x$ axis (logarithmic distance) and $0$ to $2.5$ (Mach number along $y$) along y axis. $f$ and $\alpha_\pi$ are
written on the left and upper axis respectively. All the other parameters are the same: the location of the
sonic point $x_{in} =2.795$, and the angular momentum at the inner edge $l_{in}=1.65$. $\gamma=4/3$. $f=1$ solutions
are the correct ADAF solutions. Unlike Narayan and collaborators model, our solutions show that ADAF
(indeed all the solutions with $f>0.5$ in this particular set of parameters)
solutions exist only for small viscosity. There are no shocks.
This is also confirmed by works of Bisnovatyi-Kogan (this volume) who shows that ADAF solutions for
$f>0.75$ are not possible. Non-ADAF solutions ($f \neq 1$, especially $f\leq 0.5$) are more promising, however. We note that 
there are two critical viscosities at which the topology changes dramatically (this behaviour of topologies
revolutionized our understanding of the accretion flows in black holes). For very low viscosities 
the flow passes through the inner sonic point and can join a Keplerian flow (subsonic branch with $l=l_K$)
very far away. For very high viscosities, the subsonic branch touches a Keplerian disk near by. For the
intermediate viscosities, topologies are semi-closed, but they can be reached using the outer sonic point (discussed 
in C96a) when shock conditions are satisfied. If shock conditions are not satisfied then the flow
has to enter through outer sonic point only, much like the original Bondi flow.

Several authors have lamented that they could not find shocks (Chen et al, 1997; Narayan et al. 1997;
Narayan, 1997). The solutions of these groups are not correct, as they use
the slim disk approach by specifying too many parameters at the launching point in a Keplerian disk. We have
already mentioned in \S 1 that an accreting flow cannot be strictly Keplerian.
Thus, ADAF solutions published by these groups, all of them, in my view, 
are incorrect. It is therefore immaterial whether
these solutions contained shocks or not. Only mathematically correct solution is that of Narayan \& Yi
(1994) where self-similar solution was studied. But a black hole accretion is not at all self-similar.
Exact solutions of the same set of equations (Fig. 3 above, for example) show very rich behaviour,
with Mach number going up and down several times, while a self-similar flow has a constant Mach number.
Some more reasons are written by Lu (this volume).

The correct approach is to start from a sonic point and integrate backward till the Keplerian value is reached,
and forward till the horizon is reached. That way all the sonic points are used properly 
(unlike in slim disk approach where at most one sonic point may be used, if integrated properly). 
There are a large number of independent groups (Lu et al., 1996; Yang \& Kafatos, 1994; Nobuta \& Hanawa, 1994 etc.)
who have found shocks in accretion flows. Shock study was also common in
winds from stars (e.g., Habbal \& Rosner, 1984, and references therein). While shocks and centrifugal
barrier supported boundary layer (CENBOL) would remain the most important ingredients in accretion flows close to a black hole,
the problem would still remain at the junction
point when the viscosities and cooling parameters are kept constant in the advective region.
As Chakrabarti et al. (1996) showed using a detailed numerical simulations,
at the junction the disk becomes super-Keplerian. We conjecture that either the viscosity and cooling
parameters would have to smoothly vary at the junction to  connect the advective disk with a Keplerian disk,
or, the advective disk would be a bit unsteady to try to match with a Keplerian disk. This may even 
cause quasi-periodic oscillations.

\begin{figure}
\vbox{
\vskip 0.0cm
\hskip -1.0cm
\centerline{
\psfig{figure=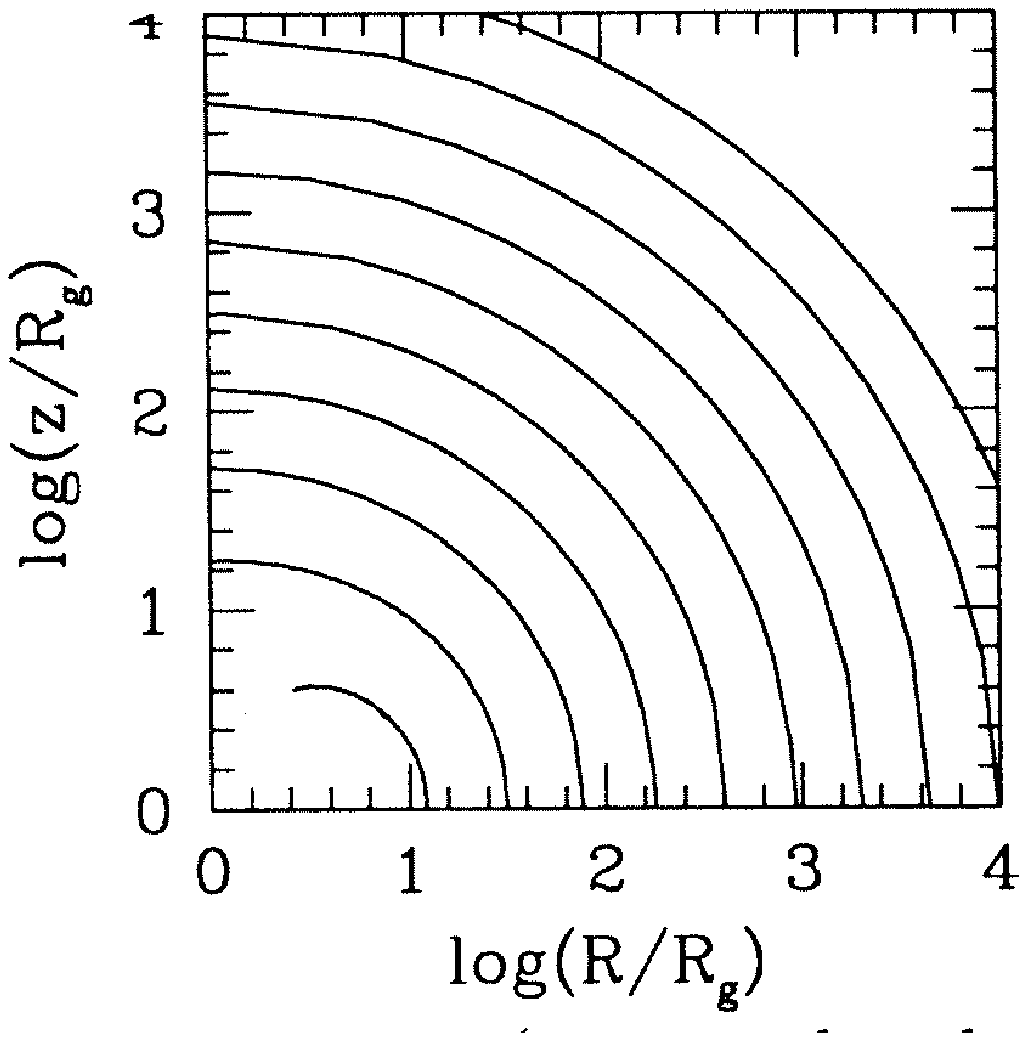,height=6truecm,width=7truecm,angle=0}}}
\vspace{-0.0cm}
\noindent {\small {\bf Fig. 4a} : Isodensity contours in an ADAF solution. Rotating flow
also moves in on the jet axis without forming any centrifugal barrier (taken from Narayan, 1997;
courtesy of the Astronomical Society of Pacific Conference Series).}
\end{figure}
\begin{figure}
\vbox{
\vskip -2.0cm
\hskip 0.0cm
\centerline{
\psfig{figure=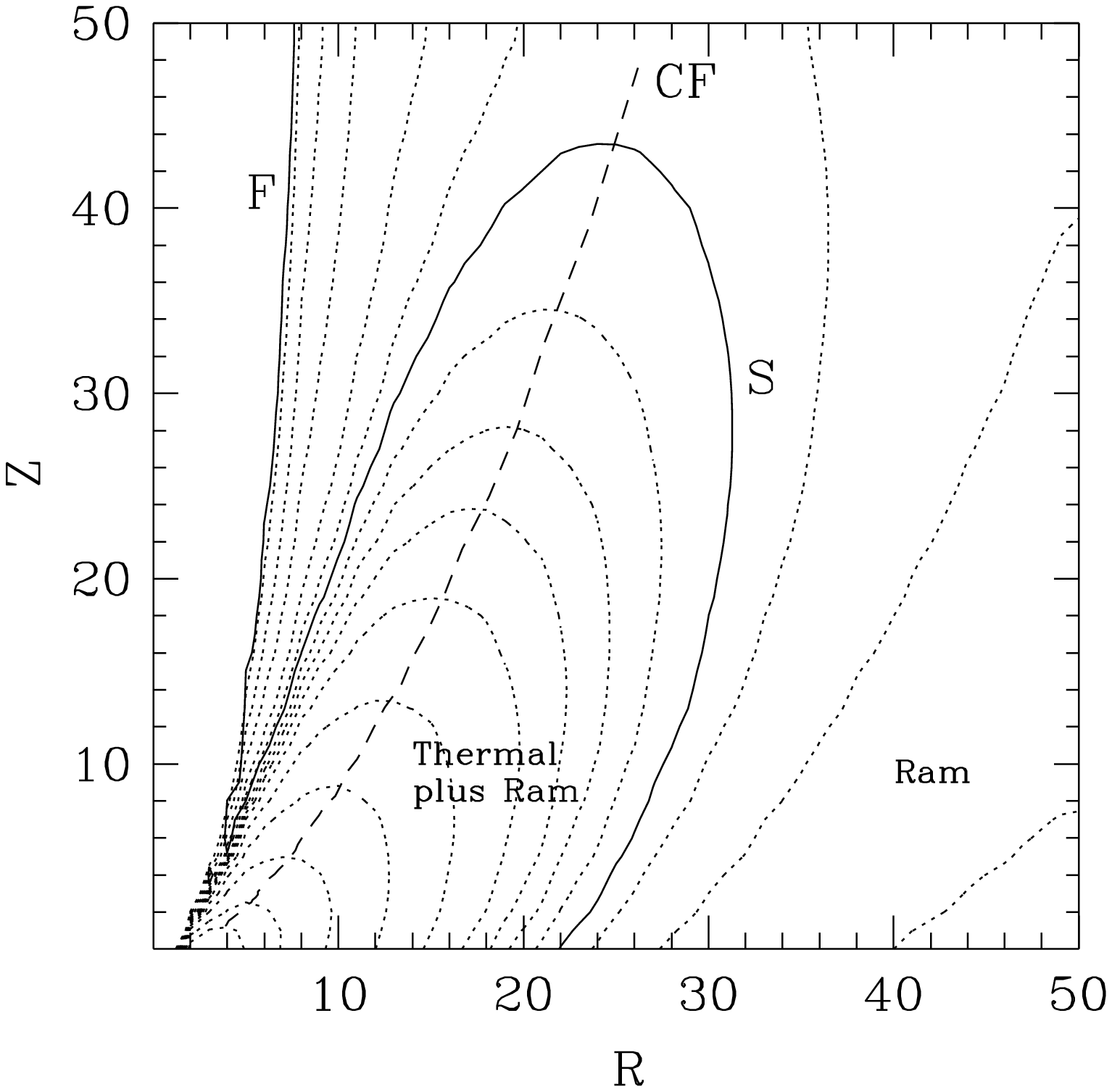,height=7truecm,width=7truecm,angle=0}}}
\vspace{-1.0cm}
\noindent {\small {\bf Fig. 4b} : Isobaric (isodensity for polytropic flows)
contours for advective flows which have centrifugal barrier CF, 
funnel wall F and possible shocks S (C96c). }
\end{figure}
We end this Section with a comparison of a `solution' of ADAF and the solution of an advective disk.
Figure 4a is supposed to be a two dimensional flow density contours of ADAF (Narayan, 1997)
and Fig. 4b shows contours of thermal pressure inside the shock (marked S) and thermal plus ram
pressure outside the shock (taken from C96c). 
In ADAF `solution', even when the matter has angular momentum, it 
has no problem on the axis. This is fundamentally incorrect. In Fig. 4b, the funnel wall (F)
and the centrifugal barrier (CF) are formed as
expected. If, for the sake of argument one assumed that in ADAF all angular momentum is removed
completely, from Fig. 2, one notes that there is no solution  close to a black 
hole which simultaneously has zero angular momentum and at the same time has negative energy
(coming from Keplerian disk). These does not imply that the original philosophy of Rees et al. (1982)
is incorrect. Disks with very low accretion rate might be possible, only if the
magnetic heating were negligible (see, Bisnovatyi-Kogan, this volume).

\section{What Should a Realistic Accretion Disk Look Like?}

\begin{figure}
\vbox{
\vskip -5.5cm
\hskip 6.0cm
\centerline{
\psfig{figure=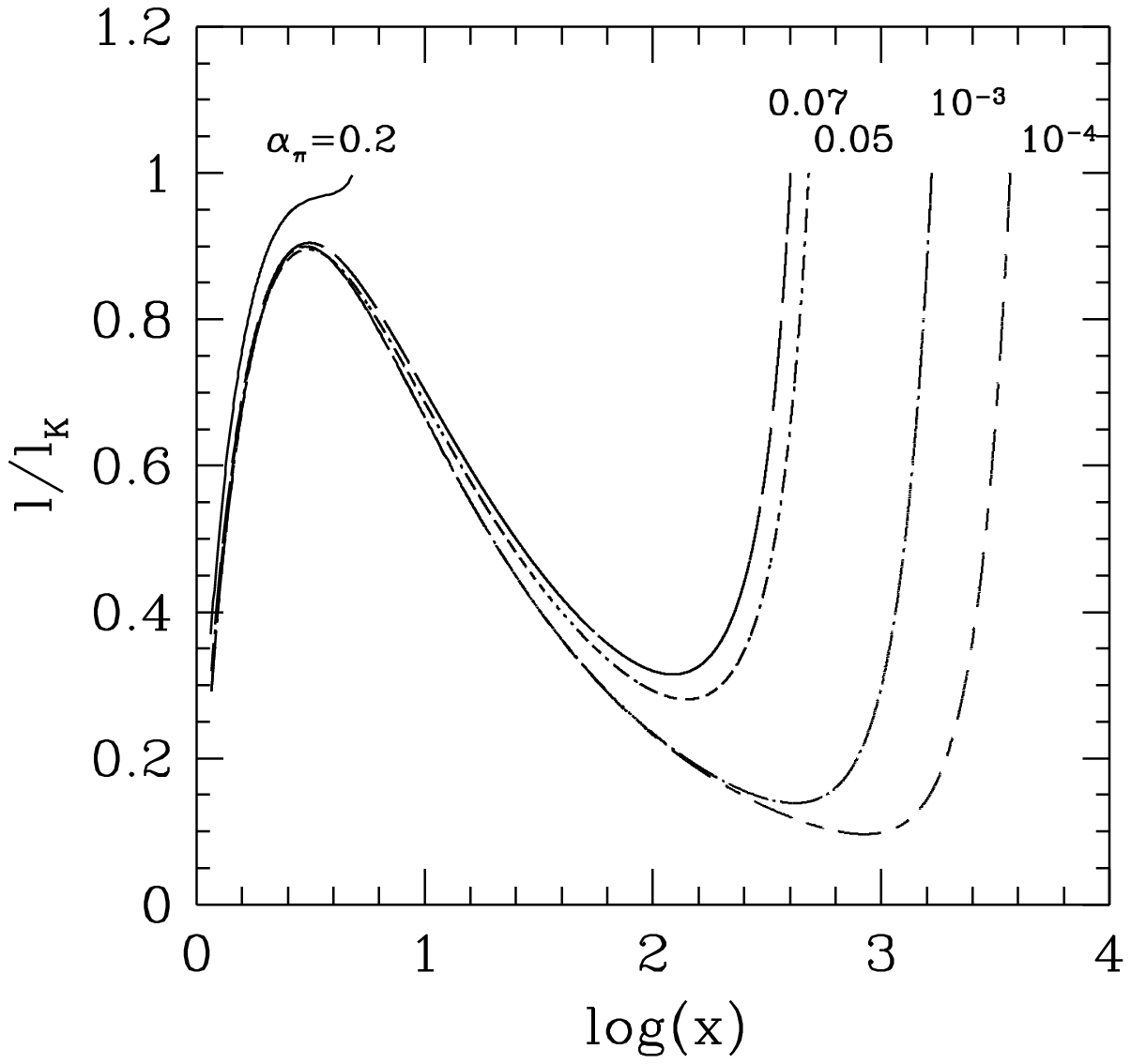,height=12truecm,width=12truecm,angle=0}}}
\vspace{-0.0cm}
\noindent {\small {\bf Fig. 5} : Ratio of disk angular momentum to the Keplerian angular momentum
of several advective disk solutions whose viscosity parameters ($\alpha_\Pi$) are marked. At or near $l/l_K=1$,
the disk may join with a Keplerian flow if energetics are right. 
Note that size of the advective region inversely varies with $\alpha_\Pi$.}  
\end{figure}
Figure 5 shows the angular momentum distribution of a collection of
solutions with $\gamma=4/3$ for different choices of the viscosity parameter
$\alpha_\Pi$ (marked on each curve). Each distribution touches a location $x_K$
where $l/l_K=1$, where, roughly speaking, one would expect the advective
region to join a Keplerian disk. First note that when other parameters (basically, 
specific angular momentum and the location of the inner sonic point) remain roughly the same,
$x_{K}$ changes inversely with $\alpha_\Pi$. 
If one assumes, as CT95 and C97a did, that alpha viscosity parameter {\it decreases}
with vertical height, then it is clear from the general behaviour of Fig. 5
above that $x_K$ would go up with height. The disk will then look like a sandwitch
with higher viscosity matter flowing along the equatorial plane with Keplerian
disk closest to the black hole. This fact that the inner edge of the disk
should move in and out when the black hole goes in soft or hard state
(e.g., Gilfanov, Churazov \& Sunyaev, 1997; Zhang et al., 1997) are
thus naturally established from this advective disk solution. Note
the presence of two groups of solutions, one with a small viscosity, and the other with a 
large viscosity. These two groups correspond to two sets of solutions
presented in Fig. 3. In the intermediate viscosity, if the shock condition 
is not satisfied, and there is no solution which passes through the
outer sonic point, there is no continuous solution connecting the Keplerian 
disk with horizon. In that case, the only possible alternative is
to produce an unsteady flow.

\begin{figure}
\vbox{
\vskip -4.5cm
\hskip -0.8cm
\centerline{
\psfig{figure=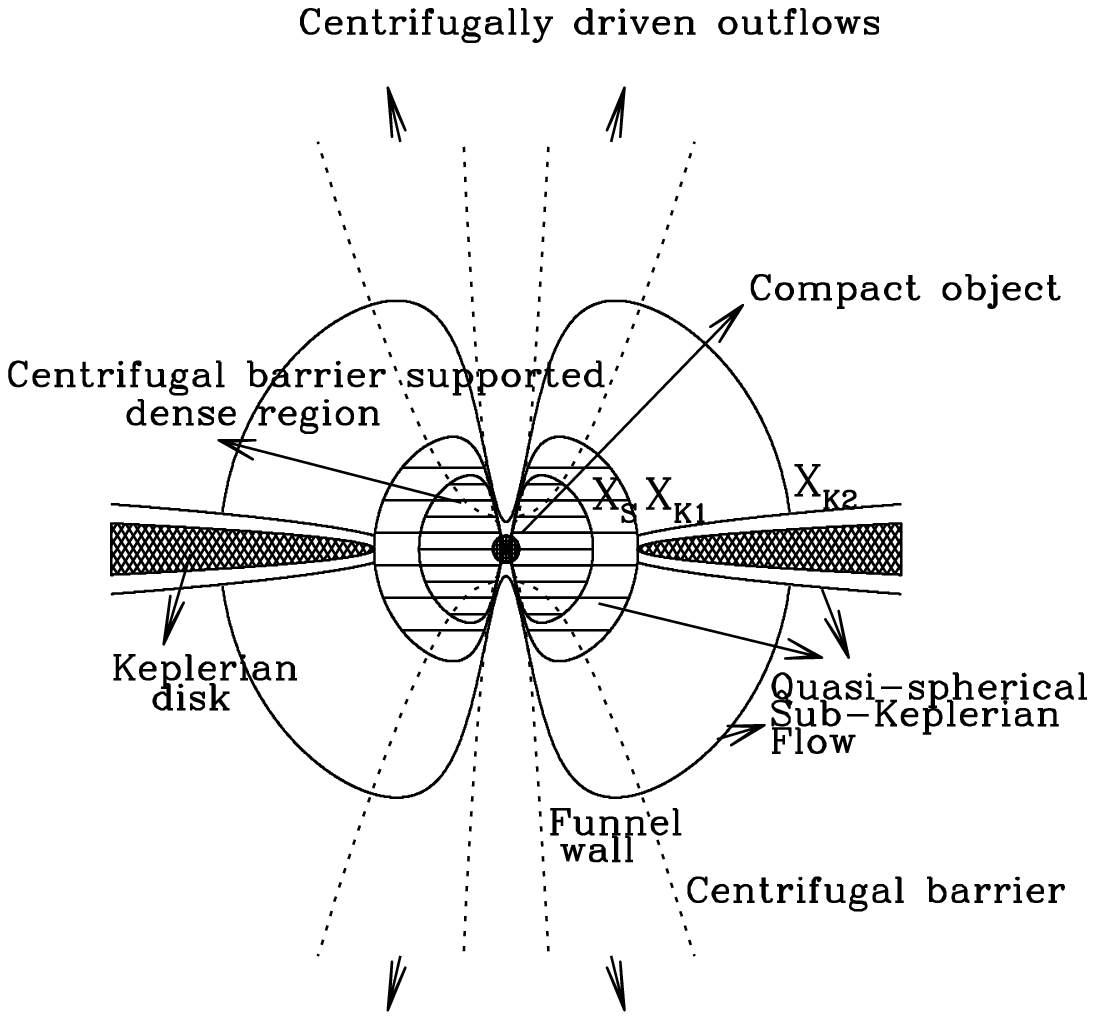,height=17truecm,width=15truecm,angle=0}}}
\vspace{-4.0cm}
\noindent {\small {\bf Fig. 6} : Schematic diagram of an unmagnetized advective disk
with all its components. Compact object is surrounded by a centrifugal barrier
supported boundary layer (CENBOL), which in turn is surrounded by an advective region.
The entire  flow may emerge from a Keplerian disk or from a mixture of Keplerian
and a sub-Keplerian flow. Jets and outflows are originated from the  CENBOL.}  
\end{figure}
Figure 6 shows the schematic diagram of a disk as we perceive it close to a black hole. There is,
as yet, {\it no single global solution} for the whole system. What we now
have are the bits and pieces of the behaviours of thin flows under various
parametric variations. After combining them, the whole picture emerges. This was
first presented in Texas Symposium in 1994 (Chakrabarti, 1995), and later was
used to compute spectra of black hole candidates (CT95, C97a). If the inflow parameters
are such that shocks form, then $x_s$ is the shock boundary, otherwise
it is simply the boundary of the centrifugal barrier when pressure 
effects are included. Without the pressure effects one could have
the funnel wall and the centrifugal barrier (dotted curves) in between which outflows
are likely to emerge.

In Section 6, we shall establish that the general observational results
agree with such a picture, even when the spectrum is non-stationary. 

\section{Progresses in Numerical simulation works}

Last two decades saw tremendous progress in numerical simulation work,
mostly due to the advent of faster computers with larger memory and due to better numerical algorithms. 
In 1978, Wilson showed that an accretion 
with significant angular momentum was accompanied by shock waves 
which traveled outwards. This code was later improved
upon, with number of grid points as well as the evolution time
orders of magnitude higher. A series of very important simulations were made
with this code to show that thick accretion disks can indeed form in inviscid
flows (Hawley, Smarr \& Wilson, 1984, 1985). These simulations also confirm the results of Wilson (1978)
that non-steady shock waves are formed which travel outward.
From the post-shock flow, a very strong wind is generated which is hollow in
nature which `hugs' the funnel wall. Due to inviscid nature of the flow,
centrifugal force kept it away from the axis of symmetry.

However, there was one problem: since in the contemporary period, only available
theoretical work on non-Keplerian disk model was that of a thick disk, the numerical
results of the thermodynamic quantities could not be compared properly. Figures 7(a-b)
show examples of two numerical solutions of Hawley (1984) which are compared with
thick disk solutions. There is only {\it qualitative} similarity between the 
theoretical work and the numerical work.  This discrepancy was mostly due to the
fact that the thick disk model was not {\it advective} while the numerical simulation
allowed the matter to rush to the black hole as fast it wanted! Of course, to a very smaller
extent, the discrepancy was due to numerical error, because of numerical diffusion
of energy and angular momentum, but it could be ignored in the present discussion.
\begin{figure}
\vbox{
\vskip 0.0cm
\hskip 0.0cm
\centerline{
\psfig{figure=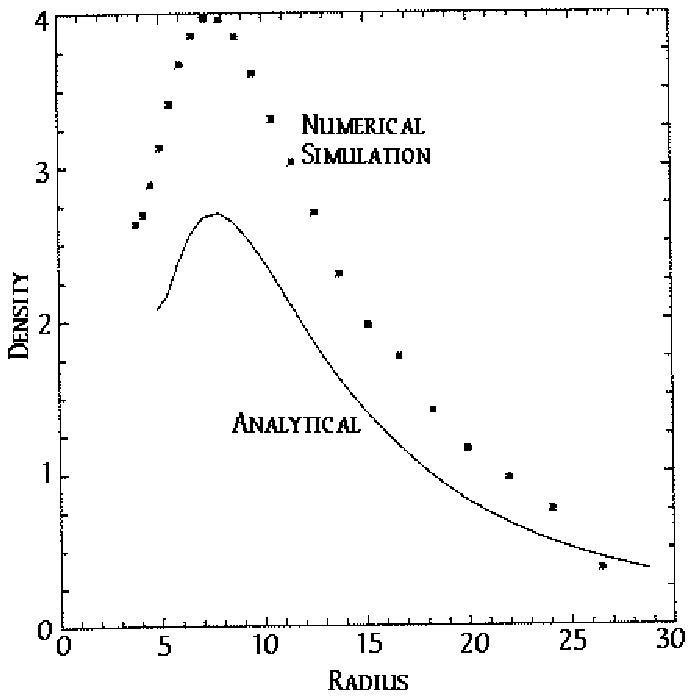,height=7truecm,width=7truecm,angle=0}}}
\vspace{-0.0cm}
\noindent {\small {\bf Fig. 7a} : Numerical results of density variation
along the equatorial plane of an inviscid flow of specific angular momentum $l=3.77GM/c$
is compared with the analytical solution from thick accretion disk. Agreement is only qualitative.
(Adapted from Hawley, 1984).  }  
\end{figure}
\begin{figure}
\vbox{
\vskip -0.5cm
\hskip 0.0cm
\centerline{
\psfig{figure=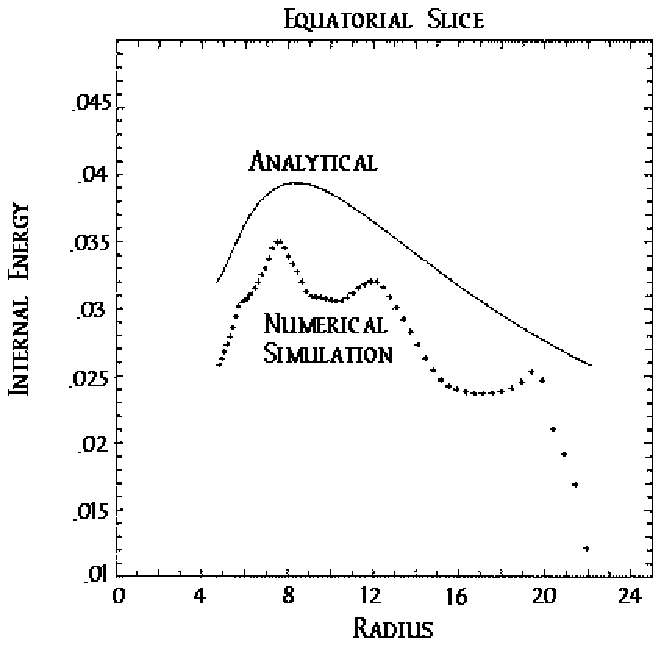,height=7truecm,width=7truecm,angle=0}}}
\vspace{-0.0cm}
\noindent {\small {\bf Fig. 7b} : Numerical results of internal energy (a measure
of temperature) variation
along the equatorial plane of an inviscid flow of specific angular momentum $l=3.8GM/c$
is compared with the analytical solution from thick accretion disk. Details do not agree.
(Adapted from Hawley, 1984). }  
\end{figure}
Today, we have a complete theoretical solution on advective disks, and a large number of 
numerical simulations (Chakrabarti \& Molteni, 1993; Molteni, Lanzafame \& Chakrabarti, 1994;
Molteni, Ryu \& Chakrabarti, 1995; see also, Ryu, this volume; Molteni, this volume) 
show how accurately the numerical simulation results match with the theoretical solution.
Indeed, the analytical works could now be used to test a code in a spherical coordinates.
The progress in this field is clearly obvious. 

A new understanding has emerged recently
regarding the time-dependent behaviour of the accretion flows. First, 
Chakrabarti \& Molteni (1995) showed  how a highly viscous flow can re-distribute
angular momentum inside a flow to form a Keplerian disk even when the 
inflow at the outer boundary is highly sub-Keplerian. Second, in 
some regions of the parameter space, particularly in $NSA$
and $NSW$ of Fig. 2, though steady solutions which pass only though the outer
sonic points were predicted, the flow chose to produce
shocks which started oscillating with frequencies similar to that observed in 
quasi-periodic oscillations. This is because the solution topology (Fig. 1)
has an inner sonic point through which higher entropy flow enters into the black hole. 
Since no stable shock was predicted, unsteady shocks are produced 
(see, Ryu, this volume).
On the other hand, in presence of cooling effects, shocks oscillate (even when a 
steady solution is predicted) when the cooling time scale and the infall time scale
roughly match (see, Molteni, this volume). Whereas the first type of
oscillation weakly depends on the accretion rates, 
especially when the rate is very low (i.e., even in the absence of cooling), 
the second type of oscillation strongly depends on cooling
effects, and therefore accretion rates. Because such oscillations not only 
explain the frequencies, but also the amplitude and other timing properties
(e.g., Paul, this volume), it is a good candidate for the explanation of 
the quasi-periodic oscillations observed in the black hole candidates.

\section{Progresses in Modeling Observational Results Which Required Accretion Disks}

One can list a large number of interesting spectral features 
that are observed from a black hole candidate (see, Chakrabarti, 1998 for a review).
The spectrum could be hard or soft or it could have a power law hard tail even in soft states
(Tanaka \& Lewin, 1995) or it could show X-ray novae  behaviour where different components have 
completely different timing properties (Orosz et al., 1997) or it could show quasi-periodic oscillation
(Dotani, this volume) or it could be in a quiescent state for years (McClintock, Horne \& Remillard, 1995) or etc.
We discuss them briefly to appreciate how they demanded the deviation from a standard Keplerian disk of SS73 and NT73.

Black holes are being fundamentally black, their proper identifications
must necessarily include quantification of very special spectral 
signatures of radiating matter entering in them. Since the inner boundary 
condition of the flow must be unique, advective disk solutions (which 
achieve this boundary value automatically) are the only disk solutions, which,
predict spectral features most self-consistently. A single global
solution for a multidimensional flow is still missing and some details
such as magnetic field, have not been incorporated self-consistently
yet. Thus, the discussion on the spectral properties can only be a bit qualitative. This 
is easily compensated for by the support of a well developed theory which satisfactorily
explains stationary and non-stationary features around galactic as well as 
extra-galactic black holes using a single framework.

\subsection {Triggering of Hard and Soft states}

Black holes are known to show hard and soft states (see, Ebisawa, Titarchuk \& Chakrabarti, 1996; hereafter ETC96).
When the viscosity of the inflow changes, the Keplerian and sub-Keplerian components (Fig. 6) redistribute
matter among themselves depending on viscosity of the flow which at the
same time, also change the inner-edge of the Keplerian component. Sudden rise
in viscosity would bring more matter to the Keplerian component (with rate 
${\dot m}_d$) and bring the Keplerian edge closer to the black hole
(see, Chakrabarti \& Molteni, 1995 for numerical simulations) and sudden fall of viscosity
would bring more matter to the sub-Keplerian halo component (with rate ${\dot m}_h$) 
and the Keplerian component would go farther out. Disk component ${\dot m}_d$ not only 
governs the soft X-ray intensity directly coming to the observer, it also 
provides soft photons to be inverse Comptonized by sub-Keplerian CENBOL 
electrons. The CENBOL (comprised of matter coming from ${\dot m}_d$ and 
${\dot m}_h$) will remain hot and emit power law (energy spectral index, 
$F_\nu \sim \nu^{-\alpha}$, $\alpha \sim 0.5-0.7$) hard X-rays only when
its intercepted soft photons from the Keplerian disk
are insufficient, i.e., when ${\dot m}_d <<1$ to ${\dot m}_d \sim 0.1$ 
or so, while ${\dot m}_h$ is much higher. For ${\dot m}_d \sim 
0.1-0.5$ (with ${\dot m}_h \sim 1$), CENBOL cools catastrophically and no
power law is seen (this is sometimes called a high state). 
With somewhat larger ${\dot m}_d$,  the power law due to the bulk motion of 
electrons (CT95; Titarchuk, Mastichiadis \& Kylafis, 1997)
is formed at around $\alpha \sim 1.5$ (this is sometimes called 
a very high state). Such hard/soft transitions are regularly seen in black hole
candidates (Dolan et al, 1979; Ebisawa et al., 1994; Zhang et al., 1997). This $\alpha$ may weakly depend
on the flow angular momentum (Chakrabarti, Titarchuk, Kazanas \& Ebisawa, 1996)

\subsection{Constancy of Slopes in Hard and Soft States}

Observations indicate that in hard states, power law slopes remain
almost constant with luminosity
(e.g. Sunyaev et al., 1994; Ebisawa et al., 1996; CT95; Kuznetsov et al., 1997;
Grove et al. 1998). 
Advective disks also show this property (CT95; C97a).
Particularly important is the weak power law in the soft state 
as this is not observed in neutron star candidates. 
In advective disks, matter behave democratically ($V=1$) close to a 
horizon independent of its history. This unique fact produces
unique spectra through bulk motion Comptonization and 
readily explains the weak power law tail in the soft states
(CT95; Titarchuk, Mastichiadis \& Kylafis, 1997). 
Since this part of the spectra is universal, it should have been seen even in hard
states, had the dominant spectra due to thermal Comptonization been somehow subtracted.
In the intermediate states both the power laws (hard and soft) are seen (Ling et al., 1997).

\subsection{Variation of Inner Edge of the Keplerian Component}

Observations indicate that the Keplerian disk component varies
with accretion rates (e.g. Gilfanov, Churazov \& Sunyaev, 1997). 
In advection dominated models of Narayan \& Yi (1994, 1995) such variations are achieved by 
evaporation and condensation of the disks by unknown fundamental physics. However,
such  variation is a natural property of the advective disks (Fig. 5).
As the viscosity increases, $x_{K}$ becomes smaller in viscous
time scale, at the same time more matter is added to the Keplerian
component. 

\subsection{Rise and Fall of X-ray Novae}

X-rays novae (e.g., A0620-00, GS2000+25, GS1124-68, V404 Cygni 
etc.)  produce bursts of intense X-rays which decay with time 
(decay time is typically 30d). This phenomenon may be repeated 
every tens to hundreds of years. While in persistent black 
hole candidates (such as, Cyg X-1, LMC X-1, LMC X-3) Keplerian 
and sub-Keplerian matter may partially redistribute to change 
states, in X-ray novae candidates the net 
mass accretion rate may indeed decrease with time after the outburst, even 
if some redistribution may actually take place. First qualitative 
explanation of the change of states in X-ray novae in terms of the 
advective disk model was put forward by ETC96. The biggest
advantage of the advective solution is that it automatically 
moves the inner edge of the Keplerian disk as viscosity is varied.
Similar to the dwarf novae outbursts, where the Keplerian disk 
instability is triggered far away (e.g., Cannizzo, 1993)
here also the instability may develop and cause the viscosity
to increase, and the resulting Keplerian disk with higher
accretion rate moves forward. In Fig. 8a we show the spectral evolution
of a two component advective disk whose inner edge (marked on each curve)
is approaching towards the black hole due to rise in viscosity. The component
accretion rates have been kept fixed at ${\dot m}_d=0.01$ and ${\dot m}_h=1.0$
respectively. No shock is assumed but the centrifugal barrier has a
similar effect (C97a). Here photon numbers are plotted against their energy.
Assuming that this evolution is the cause of the spectral variation
in GRO J1655-40 as reported by Orosz et al. (1997) and Haswell (this volume), we can have an idea of
viscosity working in that disk. In Fig. 8b, upper panel, we show the variation of the
photon rate with time (in arbitrary units). Along X-axis, we plotted $R^{3/2}$ (where $R$
is the inner edge of the Keplerian disk in units of $R_g$) which is a measure
of infall time $R/v(R)$. If the viscosity is 
such that it causes four days of delay between 
the soft X-rays and optical (as reported in Orosz et al. 1997)
then it seems that whole of the rising phase takes around nine days as reported.
Note also that the slope of B waveband is higher as compared to that of I as expected.
Here I, R \& B are chosen to be in energy bands of $1.24-2.2$eV, $2.2-3.16$eV and $3.16-4.13$eV respectively,
and soft and hard X-rays are in $2-12$keV and  $12-1000$KeV ranges respectively. 
The lower panel shows the hardness ratio variation with time. First the optical band rises
keeping  hard to soft almost fixed. This is marked as the {\it horizontal} branch. Then along the {\it diagonal} branch 
the hard component rises faster than the soft. Finally, towards the end of the novae rising phase, the
soft component rises rapidly, keeping the optical ratio nearly  constant. This is marked as the 
{\it vertical} component. Such detailed predictions should be verifiable with observations. Small modifications
or parameters, such as the angular momentum of the advective region, and the variation of the
accretion rate in the two components may be necessary when actually fitting the data.
\begin{figure}
\vbox{
\vskip -3.6cm
\hskip 0.0cm
\centerline{
\psfig{figure=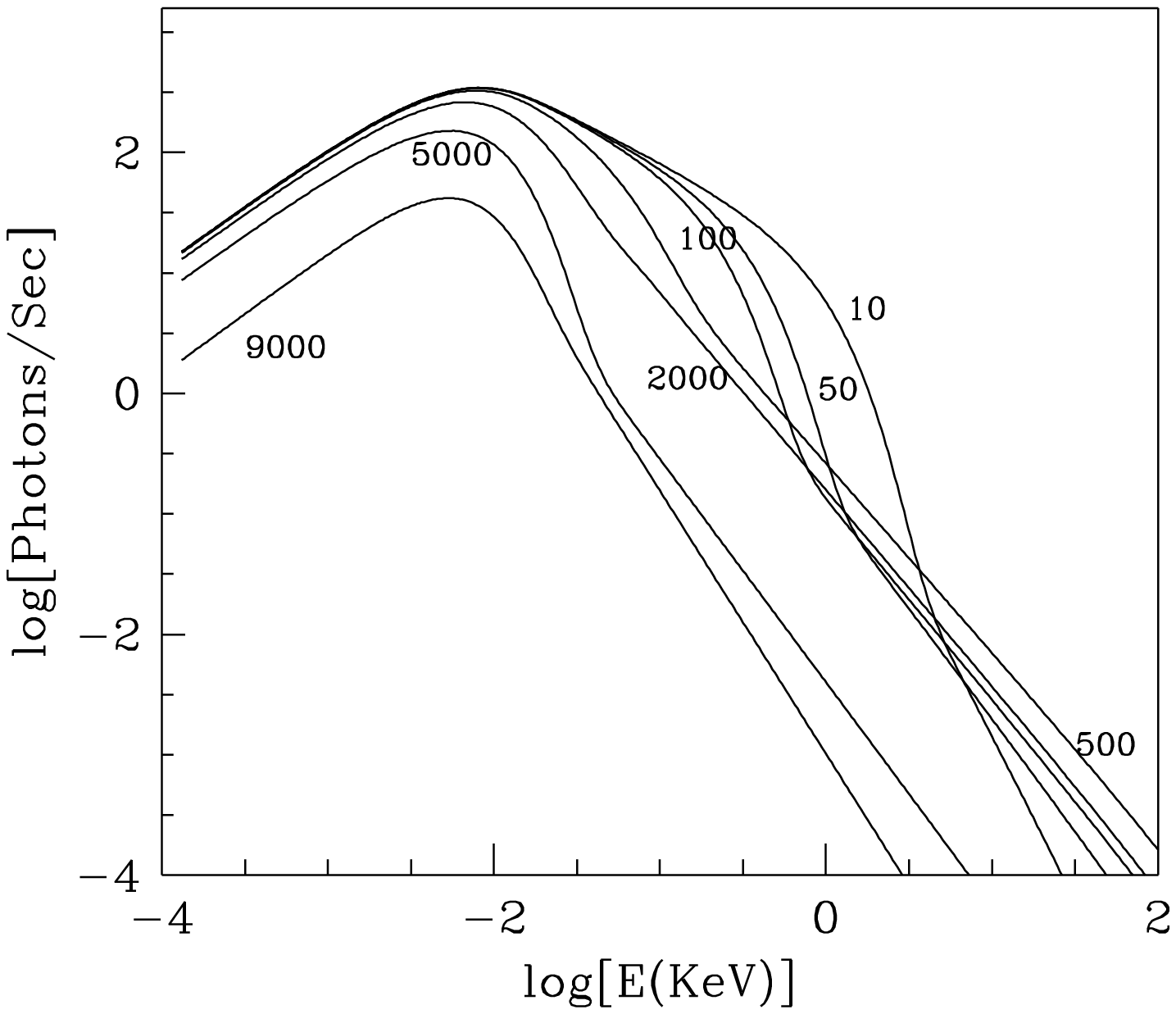,height=12truecm,width=12truecm,angle=0}}}
\vspace{-1.8cm}
\noindent {\small {\bf Fig. 8a} : Photon numbers are plotted against photon energy. 
Location of the inner edge of the Keplerian component of the disk is marked on each curve.  }  
\end{figure}
\begin{figure}
\vbox{
\vskip 0.0cm
\hskip 0.0cm
\centerline{
\psfig{figure=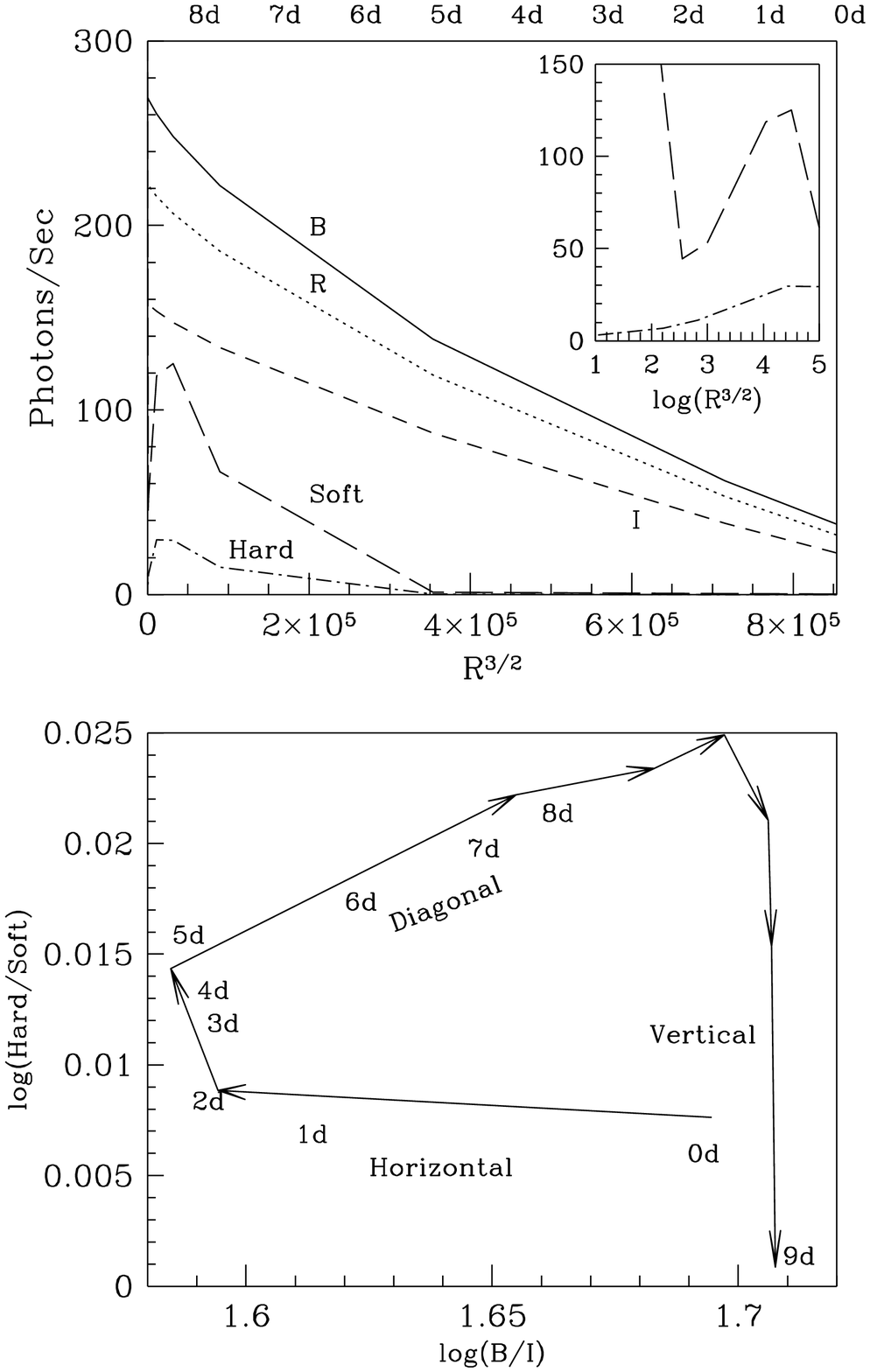,height=12truecm,width=10truecm,angle=0}}}
\vspace{-0.0cm}
\noindent {\small {\bf Fig. 8b} : Upper panel shows the rising phase of an X-ray novae while the
lower panel shows the hardness ratio. Photon numbers are plotted against $R^{3/2}$, 
a measure of infall time $R/v(R)$. Upper axis gives the number of days that are passed. Inset: details
in the last day of the rising phase. Note that rise of B band is sharper than the rise of I band; 
and soft X-ray rise is delayed by four days. In lower panel, one sees three distinct phases:
horizontal, diagonal and vertical branches. }  
\end{figure}

\subsection{Quiescent States of X-Ray Novae Candidates}

After years of X-ray bursts, the novae becomes very faint
and hardly detectable in X-rays. This is called the
quiescent state. This property
is built into advective disk models. As already 
demonstrated (Fig. 5) $x_{K}$ recedes from the black 
holes as viscosity is decreased. With the decrease of
viscosity, less matter goes to the Keplerian component (Chakrabarti \& Molteni,
1995) i.e., ${\dot m}_d$ goes down. Since the 
inner edge of the Keplerian disk does not go all the way to the
last stable orbit, optical radiation is weaker in comparison
with what it would have been predicted by a SS73
model. This behaviour is seen in V404 Cyg (Wagner et al. 1994) and A0620-00 (McClintock et al., 1995).
The deviated component from the Keplerian disk almost resembles a constant 
energy rotating flow described in detail in C89. It is also 
possible that our own galactic center may have this low viscosity, 
low accretion rate with almost zero emission efficiency 
global advective disks as mentioned in C96d.

\begin{figure}
\vbox{
\vskip 0.0cm
\hskip 0.0cm
\centerline{
\psfig{figure=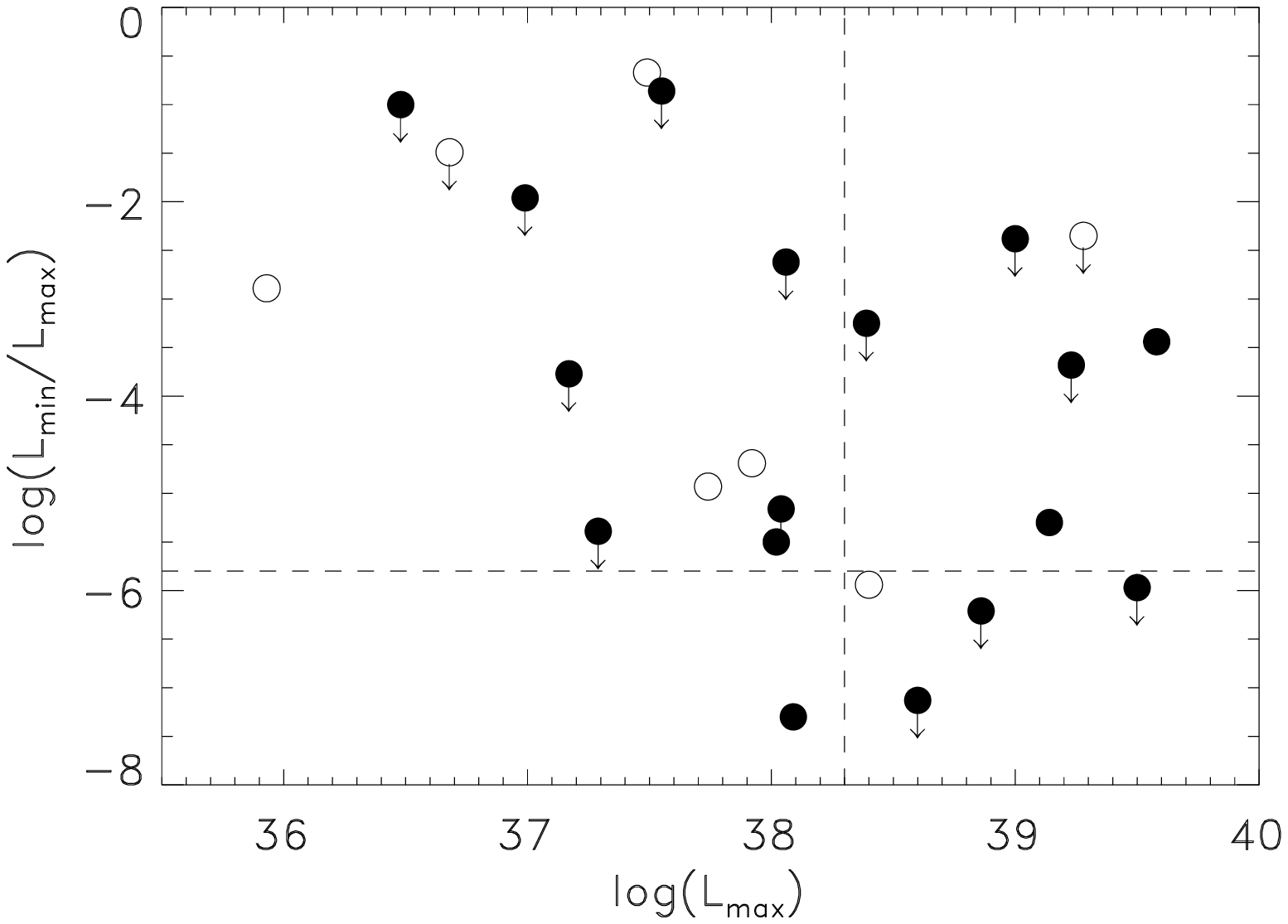,height=6.5truecm,width=8truecm,angle=0}}}
\vspace{-0.0cm}
\noindent {\small {\bf Fig. 9} : Ratio of minimum and maximum luminosities from a
collection of compact objects believed to be black holes (filled circles) and neutron stars (open circles)
[taken from Chen et al 1998]. ADAF predicts that black holes should lie below the dashed line
while neutron stars should be above it. }  
\end{figure}
Workers of ADAF `solution' claims  to fit these states (Narayan et al., 1997) well.
In this model highly viscous ($\alpha \sim 0.1-0.5$) quasi-spherical 
flow resulted from Keplerian disk evaporation (which is also in 
equipartition with magnetic field at all radii!) was used. 
Typically one data point in the hard X-ray region was used
and the fits are poor. On the contrary, the advective disk solution  we advocate
does not require such evaporation, and the advective ion torus 
of low mass accretion rate comes most naturally out of 
the governing equations only for very low viscosity case (Fig. 5).

While we are dealing with low luminosity phase of black hole candidates,
we may as well mention a recent observation by a team of experts
(Chen et al., 1998, also see, Chen, Shrader \& Livio, 1997).
They find that it is not possible to distinguish between a black hole
and a neutron star purely based on observations of total luminosity.
Figure 9 (taken from Chen et al. 1998) shows the ratio of minimum and
maximum luminosities of a collection of neutron stars (open circles)
and black holes (filled circles). The result goes against the prediction 
by the ADAF solution that all the black holes should lie below dashed line
and all the neutron stars should lie above (Garcia, McClintock \& Narayan, 1997).

\subsection{Quasi-Periodic Oscillations of X-rays}

In some large region of the parameter space
the solutions of the governing equations 1(a-d) are inherently
time-dependent. Just as a pendulum inherently oscillates, the
physical quantities of the advective disks also show oscillations of the
CENBOL region for some range in parameter space.
This oscillation is triggered by competitions 
among various time scales (such as infall time scale, cooling 
time scales by different processes). Thus, even if black holes
do not have hard surfaces, quasi-periodic oscillations could
be produced. Although any number of physical processes such as
acoustic oscillations (Taam, Chen \& Swank, 1997), 
disko-seismology (Nowak \& Wagoner, 1993), 
trapped oscillations (Kato, Honma \& Matsumoto, 1988) 
could produce such oscillation frequencies, modulation of $10-100$ 
per cent or above cannot be achieved without bringing in the dynamical 
participation of the hard X-ray emitting region, namely, the CENBOL.
By expanding back and forth (and puffing up and collapsing, alternatively)
CENBOL intercepts variable amount of soft photons
and reprocesses them. Some of the typical observational
results are presented in Dotani (this volume), Halpern \& Marshall (1996); Cui et al. (1997). 
Recently more complex behaviour has been seen in GRS 1915+105 
(Morgan, Remillard \& Greiner, 1997; Paul et al., 1998; Paul, this volume), 
which may be understood by considering  several 
cooling mechanisms simultaneously. This will be reported elsewhere.

\section{Outflows from the Advective Disks}

Problem with properly explaining observed outflows and jets
from black hole candidates is that they
have to originate not despite of the accretion flows, but because
of it. This is because unlike stellar surface, black holes 
do not have hard surfaces and atmospheres from where winds could 
independently come out. Thus physical processes must exist 
close to the black hole to join topologies of wind and accretion.
In Fig. 1 above, we see that all the wind type solutions must
have specific energy ${\cal E} \geq 1$ at the sonic point (C89). ${\cal E}$ is higher
if the flow has to be supersonic very close to the black hole (region I in Fig. 1).
In C89, it was shown that entropy measure ${\dot {\cal M}}$ must be higher
in order that flows may emerge through inner sonic point. Thus we need
to search for a physical mechanism which dumps more entropy to a little
amount of energized matter. One possible, natural source of entropy is
a stationary or non-stationary  shock wave. 

Chakrabarti (1997b) suggested one simple method to compute the outflow
rate assuming that the inflow and outflows are both conical.
Assume for the sake of argument that our system is made up of
the infalling gas, the dense boundary layer of the compact object, and the outflowing wind.
The sub-Keplerian, hot and dense, quasi-spherical region forms either due to centrifugal barrier
or due to pair plasma pressure or pre-heating effects. 
The accretion rate of the incoming flow is given by,
$$
{\dot M}_{in} = \Theta_{in} \rho \vartheta r^2 .
\eqno{(7)}
$$
Here, $\Theta_{in}$ is the solid angle subtended by the inflow, $\rho$ and
$\vartheta$ are the density and velocity respectively, and $r$ is the
radial distance in units of $GM/c^2$. In this unit, for a freely falling gas,
$$
\vartheta (r)= [\frac{1-\Gamma}{r}]^{1/2} \ \ \ \ \ {\rm and} \ \ \ \ \
\rho(r) = \frac {{\dot M}_{in}}{\Theta_{in}}(1-\Gamma)^{-1/2} r^{-3/2}
\eqno{(8)}
$$
Here, $\Gamma/r^2$ (with $\Gamma$ assumed to be a constant) is the outward radiative force.

We assume that the outer boundary of CENBOL is at $r=r_s$
where the inflow gas is compressed. The compression could be
abrupt due to  standing shock or gradual as in a shock-free flow
with angular momentum (C97a). This details are irrelevant. At this barrier, then 
$$
\rho_+(r_s) = R \rho_- (r_s)  \ \ \ \ \ {\rm and}\ \ \ \ \ \ \vartheta_+(r_s) = R^{-1} \vartheta_- (r_s) 
\eqno{(9)}
$$
where, $R$ is the compression ratio.  Exact value of the compression ratio
is a function of the flow parameters, such as the specific energy and the
angular momentum.  Here, the subscripts $-$ and $+$ denote the pre-shock and post-shock 
quantities respectively. At the shock surface, the total pressure 
(thermal pressure plus ram pressure) is balanced.
$$
P_- (r_s) + \rho_- (r_s) \vartheta_-^2 (r_s)
= P_+ (r_s) + \rho_+ (r_s) \vartheta_+^2 (r_s).
\eqno{(10)}
$$
Assuming that the thermal pressure of the pre-shock incoming flow is 
negligible compared to the ram pressure, using eqs. (10) we find,
$$
P_+(r_s) = \frac{R-1}{R} \rho_-(r_s) \vartheta_-^2 (r_s).
\eqno{(11)}
$$
The isothermal sound speed in the post-shock region is then,
$$
C_s^2= \frac{P_+}{\rho_+}=\frac{(R-1)(1-\Gamma)}{R^2}\frac{1}{r_s}
=\frac{(1-\Gamma)}{f_0 r_s}
\eqno{(12)}
$$
where, $f_0=R^2/(R-1)$. 
An outflow is expected to be subsonic close to the black hole and supersonic far away.
In the subsonic region, the pressure and density
are expected to be almost constant and thus it is customary to 
assume isothermality condition up to the sonic point (Tarafdar, 1988).
The sonic point conditions are computed from the radial momentum equation, 
$$
\vartheta \frac{d\vartheta}{dr} + \frac{1}{\rho}\frac{dP}{dr} 
+\frac{1-\Gamma}{r^2} = 0.
\eqno{(13)}
$$
and the continuity equation
$$
\frac{1}{r^2}\frac{d (\rho \vartheta r^2)}{dr} =0
\eqno{(14)}
$$
in the usual way, i.e., by eliminating $d\rho/dr$,
$$
\frac{d\vartheta}{dr}= \frac{N}{D}
\eqno{(15)}
$$
where,
$$
N=\frac{2 C_s^2}{r} - \frac{1-\Gamma}{r^2} \ \ \ \ {\rm and} \ \ \ \ D=\vartheta - \frac{C_s^2}{\vartheta}
\eqno{(16)}
$$
and putting $N=0$ and $D=0$ conditions. These conditions
yield, at the sonic point $r=r_c$, for an isothermal flow,
$$
\vartheta (r_c) = C_s,  \ \ \ \  {\rm and} \ \ \ \   
r_c = \frac{1-\Gamma}{2 C_s^2}=\frac {f_0 r_s}{2}
\eqno{(17)}
$$
where, we have utilized eq. (12) to substitute for $C_s$. 

The constancy of the integral of the radial momentum equation 
(eq. 13) in an isothermal flow gives: 
$$
C_s^2 ln \ \rho_+ -\frac{1-\Gamma}{r_s} =
\frac{1}{2}C_s^2 + C_s^2 ln \ \rho_c -\frac{1-\Gamma}{r_c}
\eqno{(18)}
$$
where, we have ignored the initial value of the outflowing 
radial velocity $\vartheta (r_s)$ at the dense region boundary, 
and also used eq. (17a). We have also put $\rho(r_c)=\rho_c$ 
and $\rho(r_s) = \rho_+$. Upon simplification, we obtain,
$$
\rho_c =\rho_+  exp (-f) \ \ \ \ {\rm where,} \ \ \ \ f= f_0 - \frac{3}{2}.
\eqno{(19)}
$$
Thus, the outflow rate is given by,
$$
{\dot M}_{out} = \Theta_{out} \rho_c \vartheta_c r_c^2 
\eqno{(20)}
$$
where, $\Theta_{out}$ is the solid angle subtended by the outflowing cone. 
Upon substitution, one obtains,
$$
\frac{{\dot M}_{out}} {{\dot M}_{in}} = R_{\dot m}
=\frac{\Theta_{out}}{\Theta_{in}} \frac{R}{4} f_0^{3/2} exp \ (-f)
\eqno{(21)}
$$
which, explicitly depends only on the compression ratio:
$$
\frac{{\dot M}_{out}}{{\dot M}_{in}} =R_{\dot m}=
\frac{\Theta_{out}}{\Theta_{in}}\frac{R}{4} 
[\frac{R^2}{R-1}]^{3/2} exp  (\frac{3}{2} - \frac{R^2}{R-1})
\eqno{(22)}
$$
apart from the geometric factors. This simple result 
is independent of the size of the dense cloud or the outward radiation 
force constant $\Gamma$. This is because the gravitational and 
radiation force have very simple forms ($\propto 1/r^2$). Also, outward driving 
centrifugal force was ignored.  Similarly, the ratio
is independent of the mass accretion rate which should be valid only for
low luminosity objects. For high luminosity flows, Comptonization would
cool the dense region completely (CT95) and the mass loss will be negligible.
In reality there would be a dependence (probably weak) on these quantities when
full general relativistic considerations of the rotating flows are 
made. Exact and detailed computations using both the transonic inflow
and outflow (where the compression ratio $R$ is also computed self-consistently)
are in Das  \& Chakrabarti (1998).

Figures 10(a-b) contain the basic results.  Figure 10a shows the
ratio $R_{\dot m}$ as a function of the compression ratio $R$ (plotted from $1$ to $7$),
and Figure 10b  shows the same quantity as a function of the
polytropic constant $n=(\gamma-1)^{-1}$ (drawn from $n=3/2$ to $3$), $\gamma$ being the adiabatic
index. In Fig. 10a, the curve is drawn for any generic compression ratio. 
and in Fig. 10b, the curve is drawn assuming the strong shock
limit only: $R=(\gamma+1)/(\gamma-1)=2n+1$. In both the cases, $\Theta_{out} \sim
\Theta_{in}$ has been assumed for simplicity. Note that if the compression 
does not take place (namely, if the denser region does not exist), then
there is no outflow in this model. Indeed for, $R=1$, the ratio $R_{\dot m}$ is zero as expected.
Since compression in shocks goes along with entropy generation, the outflows are associated with
large entropy generation.

In a relativistic inflow or for a radiation dominated inflow, $n=3$ and $\gamma=4/3$. 
In the strong shock limit, the compression ratio is $R=7$ and the ratio 
of inflow and outflow rates becomes,
$$
R_{\dot m}=0.052 \ \frac{\Theta_{out}}{\Theta_{in}}.
\eqno{(23a)}
$$
For the inflow of a mono-atomic ionized gas $n=3/2$ and $\gamma=5/3$. 
The compression ratio is $R=4$, and the ratio in this case becomes,
$$
R_{\dot m}=0.266 \ \frac{\Theta_{out}}{\Theta_{in}}.
\eqno{(23b)}
$$
Since $f_0$ is smaller for $\gamma=5/3$ case, the density at the
sonic point in the outflow is much higher 
(due to exponential dependence of density on $f_0$, see, eq. 19) 
which causes the higher outflow rate, even when the actual jump in density
in the postshock region, the location of the
sonic point and the velocity of the flow at the sonic point are much lower.
It is to be noted that generally for $\gamma >1.5$ shocks are not
expected (Chakrabarti, 1990), but the centrifugal barrier supported dense region
would still exist. As is clear, the entire behavior of the outflow
depends only on the compression ratio, $R$ and the collimating
property of the outflow $\Theta_{out}/\Theta_{in}$.

\begin{figure}
\vbox{
\vskip -5.0cm
\hskip 0.0cm
\centerline{
\psfig{figure=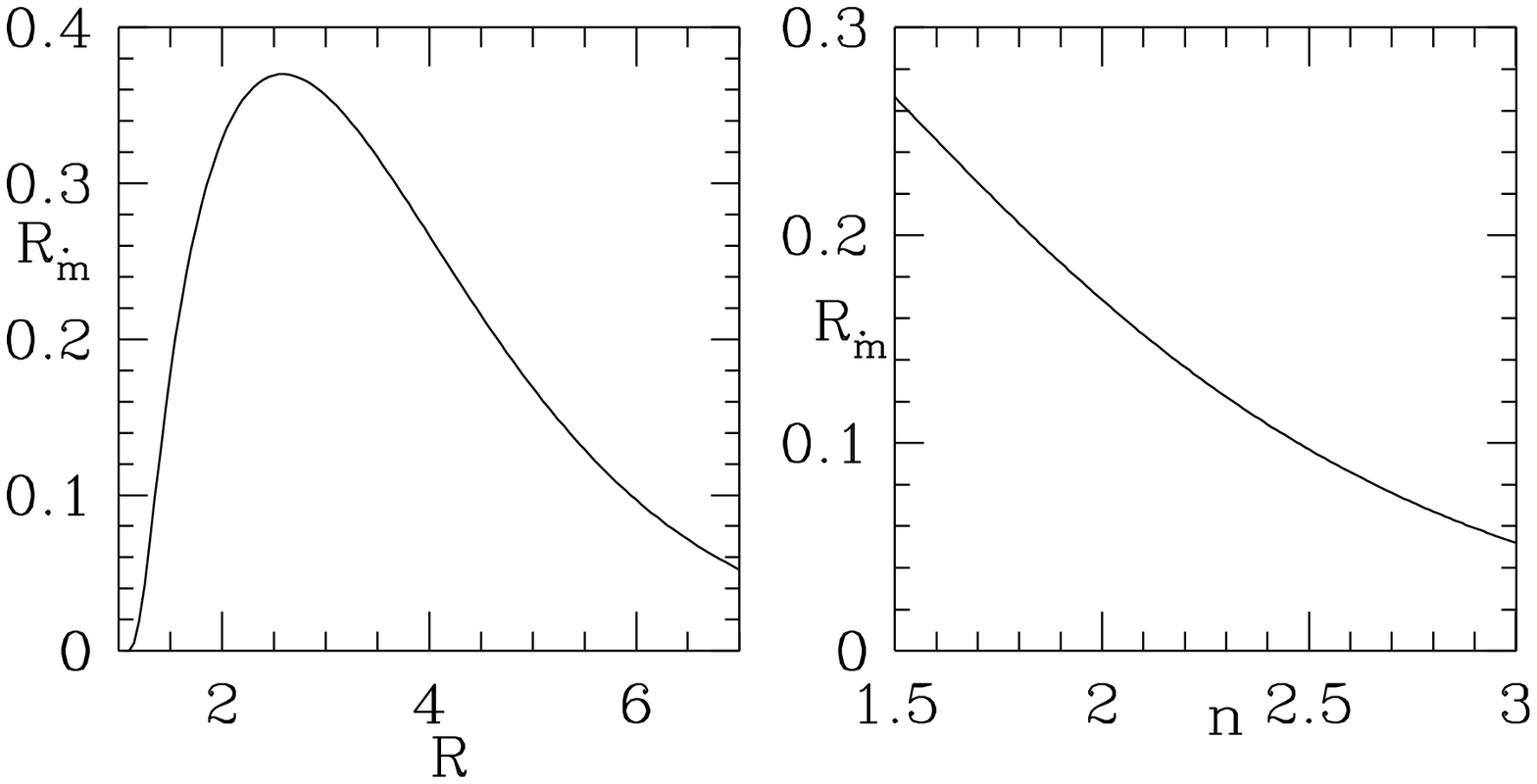,height=12truecm,width=12truecm,angle=0}}}
\vspace{-2.0cm}
\noindent {\small {\bf Fig. 10 (a-b)} : Ratio ${\dot R}_{\dot m}$ of the outflow rate and the inflow rate
as a function of the compression ratio of the gas at the dense region boundary (a)
and its variation with the polytropic
constant $n$ in the {\it strong shock limit} (b). Solid angles subtended by the inflow
and the outflow are assumed to be comparable.  }  
\end{figure}
Outflows are usually concentrated near the vertical axis, while the inflow is near
the equatorial plane. Assuming a half angle of $10^o$ in each case, 
we obtain,
$$
\Theta_{in}= \frac {2 \pi^2}{9}; \ \ \ \ \ \Theta_{out}= \frac {\pi^3}{162}; \ \ \ \frac{\Theta_{out}}{\Theta_{in}} =\frac{\pi}{36} .
\eqno{(24)}
$$
The ratios of the rates for $\gamma=4/3$ and $\gamma=5/3$ are then
$$
R_{\dot m}=0.0045 \ \ \ \ \ {\rm and} \ \ \ \ \ R_{\dot m}= 0.023
\eqno{(25)}
$$
respectively. Thus, in quasi-spherical systems, 
in the case of strong shock limit, the outflow rate is at the most a couple 
of percent of the inflow. 

It is to be noted that although the existence of outflows are well known, 
their rates are not. The only definite candidate whose outflow rate is known with 
any certainty is probably SS433 (a neutron star?) whose mass outflow rate was estimated to be
${\dot M}_{out} \gsim 1.6 \times 10^{-6}  f^{-1} n_{13}^{-1} D_5^2 M_{\odot} $ yr$^{-1}$
(Watson et al. 1986), where $f$ is the volume filling factor, $n_{13}$ 
is the electron density $n_e$ in units of $10^{13}$ cm$^{-3}$, $D_5$ 
is the distance of SS433 in units of $5$kpc. Considering a central 
object of mass $3M_{\odot}$, the Eddington rate is ${\dot M}_{Ed} \sim 
0.6 \times 10^{-8} M_{\odot} $ yr$^{-1}$ and assuming an efficiency 
of conversion of rest mass into gravitational energy $\eta \sim 0.1$, the 
critical rate would be roughly ${\dot M}_{crit} = {\dot M}_{Ed} / \eta \sim
6 \times 10^{-8} M_{\odot} $ yr$^{-1}$. Thus, in order to produce the outflow rate
mentioned above even with our highest possible estimated $R_{\dot m}\sim 0.4$ (see,
Fig. 10a), one must have ${\dot M}_{in} \sim 67 {\dot M}_{crit}$ which is very high
indeed. One possible reason why the above rate might have been over-estimated
would be that below $10^{12}$cm from the central mass (Watson et al. 1986), $n_{13} >>1 $ 
because of the existence of the dense region at the base of the outflow. 

In numerical simulations the ratio of the outflow and inflow has been computed
in several occasions (Eggum, Coroniti \& Katz, 1985; Molteni, Lanzafame \& Chakrabarti,
1994). Eggum et al. (1985)  found the ratio to be $R_{\dot m} \sim 0.004$ for a 
radiation pressure dominated flow. This is generally comparable to what we found
above (eq. 25). In Molteni et al. (1994) the centrifugally driven outflowing wind 
generated a ratio of $R_{\dot m}\sim 0.1$. Here, the angular momentum was present 
in both inflow as well as outflow, and the shock was not very strong. Thus,
the result is again comparable with what we find here. The detailed 
computations are presented in Das \& Chakrabarti (1998).

\section{Future Directions}

The advective disks brought in a lot of possibilities and we are only beginning to understand them.
Advective disks being hotter, a significant nucleosynthesis
can take place inside them. Chakrabarti, Jin and Arnett (1987) and Jin, Arnett \& Chakrabarti (1989) initiated the
studies of nucleosynthesis in black hole accretion disks. These studies were made
using `then contemporary' accretion disks, namely, thick disks. In the recent studies,
Mukhopadhyay \& Chakrabarti (1998, see also Mukhopadhyay, this volume) finds that 
nuclear composition change could be significant and what is more, disks may even show
instability as various elements are depleted in different radial distances.
The theoretical understanding of the stability of the disk in presence
of nucleosynthesis is a subject of intense study (Ray \& Chakrabarti, 1998) and 
results would be reported elsewhere. Another important direction is to see if the
two temperature character of the flow is maintained in presence of magnetic heating
as recently pointed out by Bisnovatyi-Kogan (this volume). Most certainly, decoupling
of protons and electrons as required by ADAF is not possible. 

Twenty five years have passed by to appreciate the fact that the inertial force of
matter in an accreting flow could be important to alter the basic topologies of
disk accretion. Last two decades, new disks `models'  mushroomed every so often
that it became customary to use one model in one occasion
and another model in another occasion. Lately, what people have realized we have to go back
to basics: the governing equations are unique but the solutions would depend on input parameters.
One should be more concerned about the nature of the solutions rather than ad hoc models. 
These zoo of solutions hold the key to solving diverse problems in black hole astrophysics. 
Since advection can be thought to be  synonymous with accretion, and it is no surprise 
that advective disks can resolve most of the outstanding issues very naturally. Future of this subject 
depends on understanding these new solutions.

{}
\end{document}